\RequirePackage{fix-cm}
\documentclass[twocolumn,epjc3]{svjour3}
\smartqed  
\RequirePackage{graphicx}
\RequirePackage{mathptmx}      

\usepackage{cite}
\usepackage{appendix}
\usepackage{epsfig}
\usepackage{amsmath}
\usepackage{amssymb}
\usepackage{bm}
\usepackage{hyperref}
\usepackage{slashed}
\newcommand{\be} {\begin{equation}}
\newcommand{\ee} {\end{equation}}
\newcommand{\ba} {\begin{eqnarray}}
\newcommand{\ea} {\end{eqnarray}}

\newcommand{\cpeps}{\epsilon^{\rm CP}}

\newcommand{\cL} {\mathcal L}

\newcommand{\GeV}{\text{ GeV}}
\newcommand{\SU}{\text{SU}}
\newcommand{\U}{\text{U}}

\usepackage{color}
\definecolor{darkblue}{cmyk}{1,0.3,0,0.2}
\definecolor{violet}{cmyk}{0,1,0,0.2}
\hypersetup{colorlinks, bookmarksnumbered, citecolor=darkblue, linkcolor=darkblue, pdfstartview=FitH, urlcolor=darkblue, linktocpage}

\newcommand{\arXhref}[1]{\href{http://arxiv.org/abs/#1}{#1}}

\journalname{Eur. Phys. J. C}

\begin{document}

\title{Electroweak bounds on Higgs  pseudo-observables and $h\to 4\ell$ decays}

\author{Mart\'in Gonz\'alez-Alonso\thanksref{addr1}, Admir Greljo\thanksref{addr2}, \\[0.1cm]
   Gino Isidori\thanksref{addr2,addr3}, David Marzocca\thanksref{addr2}
}

\institute{IPN de Lyon/CNRS, Universit\'e Lyon 1, Villeurbanne, France \label{addr1}
\and
Physik-Institut, Universit\"at Z\"urich, CH-8057 Z\"urich, Switzerland \label{addr2}
\and
INFN, Laboratori Nazionali di Frascati, I-00044 Frascati, Italy \label{addr3}
}

\maketitle

\begin{abstract}
We analyze the bounds on the Higgs pseudo-ob\-ser\-vables following 
from electroweak constraints, under the assumption that the Higgs particle 
is the massive excitation of an $\SU(2)_L$ doublet. Using such bounds,
detailed predictions for $h\to 4\ell$ decay rates, dilepton spectra, and lepton-universality 
ratios are presented.
\end{abstract}

\section{Introduction}
\label{intro}

The decays of the Higgs particle, $h(125)$, can be characterized by a set of pseudo-observables (PO)
that describe, in great generality, possible deviations  from the Standard Model (SM) in the limit of heavy New Physics (NP)~\cite{Gonzalez-Alonso:2014eva}.
These PO should be considered as independent variables in the absence of specific symmetry or dynamical assumptions; 
however, relations among themselves and also between Higgs and non-Higgs PO arise  in specific NP frameworks.
Testing if such relations are verified by data provides a systematic way to investigate 
the nature of the Higgs particle and, more generally, to test the underlying symmetries of physics beyond the SM.

The constraints on the Higgs PO following from the hypotheses of CP invariance, flavor universality and custodial symmetry 
have   been discussed in Ref.~\cite{Gonzalez-Alonso:2014eva}. These symmetries lead to a series of relations 
among PO that can be tested using Higgs data only. In this paper we analyze the constraints
following from the hypothesis that the Higgs particle is the massive excitation of a pure $\SU(2)_L$ doublet, 
i.e.~constraints and relations among the PO that hold in the so-called linear Effective Field Theory 
(EFT) regime. 

Under this assumption, the $h$ field appears in the effective  SM+NP Lagrangian 
through the combination $(v + h)^n$, where $v \approx 246 \GeV$ 
is the $\SU(2)_L$-breaking vacuum expectation value.  
This implies that processes involving the Higgs particle can be related to 
Electroweak (EW) precision observables that do not involve the physical Higgs boson.  
As pointed out in Ref.~\cite{Isidori:2013cga}, testing if such relations are satisfied 
represents a very powerful tool to discriminate linear and non-linear EFT approaches to Higgs physics.
In particular, sizable deviations from the SM in the $h\to 4\ell$  spectra are allowed, in general, 
in the non-linear EFT (i.e.~if $v$ and $h$ are decoupled)~\cite{Isidori:2013cla},
while they are significantly constrained by EW precision observables 
in the linear EFT~\cite{Contino:2013kra,Dumont:2013wma,Pomarol:2013zra}.
Observing sizable deviations from the SM in the $h\to 4\ell$ spectra could therefore 
allow to exclude that  $h$ is the massive excitation of a pure $\SU(2)_L$ doublet~\cite{Isidori:2013cga}
(for additional tests about the $\SU(2)_L$ properties of the $h$ boson 
see Ref.~\cite{Brivio:2013pma}).

In order to precisely quantify the above statement, in this paper we present a systematic evaluation 
of the bounds on the Higgs PO following the EW constraints in the linear EFT regime,
with particular attention to the PO entering both $h\to 4\ell$ and $h\to 2\ell2\nu$ decays. 
Using such bounds, we derive predictions for $h\to 4\ell$ decay rates, dilepton invariant-mass spectra, and lepton-universality ratios.

\section{Relating Higgs pseudo-observables to EW observables}
\label{sec:LEPandHPO}

Given the present and near-future level of precision in Higgs physics, and the absence of any significant deviation from the SM,
it is a good approximation to work at the tree level in the linear EFT, as far as NP effects are concerned.
In this limit, the Higgs PO ($\kappa_i$ and $\epsilon_i $) can be expressed as linear combinations of 
the Wilson coefficients of the EFT Lagrangian.   Analogously,   the EW PO, such as the  
$Z$- and $W$-pole effective couplings, the $W$ mass, and the effective Triple Gauge boson Couplings (TGC), can be expressed 
as linear combinations of the same Wilson coefficients. By inverting these relations it is possible to get rid of some of the (basis-de\-pen\-dent) Wilson 
coefficients and derive basis-independent relations between Higgs and EW PO. 

In doing so, one realizes that the Higgs contact terms $\epsilon_{Zf}$ and $\epsilon_{Wf}$~\cite{Gonzalez-Alonso:2014eva} can be expressed in a closed form in terms of quantities already strongly constrained by LEP and Tevatron data~\cite{Pomarol:2013zra,Gupta:2014rxa,Pomarol:2014dya}, such as the $Z$ and $W$ effective on-shell couplings to fermions, and the effective anomalous TGC:
\ba
	\epsilon_{Z f} &=& \frac{2m_Z}{v} \left( \delta g^{Zf} - (c_\theta^2 T^3_f + s_\theta^2 Y_f) {\bf 1}_3 \delta g_{1,z} + t_\theta^2 Y_f {\bf 1}_3 \delta \kappa_\gamma \right)~,
	\nonumber\\
	\epsilon_{W f} &=& \frac{\sqrt{2} m_W}{v} \left( \delta g^{Wf} - c_\theta^2 {\bf 1}_3 \delta g_{1,z}  \right)~.
	\label{eq:cont_terms_EFT}
\ea
Here, generalising the notation of Ref.~\cite{Gonzalez-Alonso:2014eva}, we treat $\epsilon_{Z f}$ and $\epsilon_{W f}$ as $3\times 3$ matrices in flavor 
space (with implicit flavor indices). On the right-hand side,  $\delta g^{Zf}$ and $\delta g^{Wf}$ denote the anomalous effective on-shell $Z$ and $W$ 
couplings to the fer\-mion $f$, again in an implicit $3\times 3$ notation,  ${\bf 1}_3$ is the identity matrix,
and $\delta g_{1,z}$ and $\delta \kappa_\gamma$ are the effective anomalous TGC extracted from 
$e^+ e^- \rightarrow W^+ W^-$ and single $W$ production~\cite{Falkowski:2014tna} (see~\ref{app:HiggsBasis} for the definition of the various terms).\footnote{We stress that the pseudo-observables $\epsilon_{W f}$ and $\delta g^{Wf}$, which in general are complex, are real in the linear EFT scenario~\cite{Gonzalez-Alonso:2014eva}.} The parameters $\{c_\theta, s_\theta, t_\theta\}$ denote the cosine, sine and tangent of the Weinberg angle, defined as in Ref.~\cite{HiggsBasis}.

The parameters $\delta g^{Zf}$ relevant to this work are the leptonic $Z$ couplings, which have been constrained at the per-mil level at LEP-I \cite{ALEPH:2005ab,Pomarol:2013zra,Falkowski:2014tna}. Per-mil constraints on the $\delta g^{Z\ell}$ hold also in the most generic flavor scenario~\cite{Efrati:2015aaa},
and the (mild) relaxation of the bounds due to off-shell $Z$ effects~\cite{Berthier:2015oma} has no practical consequences to
the present analysis. This implies that lepton flavor non-universal effects in the Z couplings are strongly suppressed. The leptonic $W$ couplings $\delta g^{W\ell}$ are instead constrained only at the percent level. In the following we consider the bounds from the non-universal fit of Ref.~\cite{Efrati:2015aaa}, reported also in \ref{app:HiggsBasis}.

In general, the parameters describing anomalous TGC in the effective Lagrangian are not PO~\cite{Trott:2014dma}.  Here we follow the approach of 
Ref.~\cite{Falkowski:2014tna} where the $e^+ e^- \rightarrow W^+ W^-$ cross-section is parameterized in terms of the effective on-shell $Z$ and $W$ 
couplings to fermions plus the three parameters $\{\delta g_{1,z},  \delta \kappa_\gamma,  \lambda_Z\}$, which therefore represent a consistent  TGC PO set. 
The constraints on this set  obtained in Ref.~\cite{Falkowski:2014tna} are collected in~\ref{app:HiggsBasis}. It should be stressed that a flat direction is present when all three TGC PO are included at the linear level \cite{Brooijmans:2014eja}
(see also~\cite{Alonso:2013hga}), 
which reflects into a very loose bound on $\delta g_{1,z}$ when $\lambda_Z$ is 
marginalized. In the following we will present results both for this case and for the case where $\lambda_Z$ is fixed to zero, which is a common condition in many interesting explicit UV models.\footnote{The flat TGC direction is lifted if quadratic terms in the cross section are included. 
In principle, this procedure is not consistent with  the EFT power counting, given the lack of inclusion of contributions from $d=8$ operators, that are formally of the same order. 
However,  in Ref.~\cite{Falkowski:2014tna} it is argued that the result of the quadratic fit are consistent with the EFT expansion since 
higher-dimension operators contributing to the $s$-channel give a suppressed contribution. For our purposes, we notice that the constraints on $\{\delta g_{1,z},  \delta \kappa_\gamma\}$
obtained from the quadratic fit are essentially equivalent to those obtained  setting $\lambda_Z = 0$.} 

\begin{figure*}[t]
  \centering
  \begin{tabular}{cc}
   \includegraphics[width=0.45\textwidth]{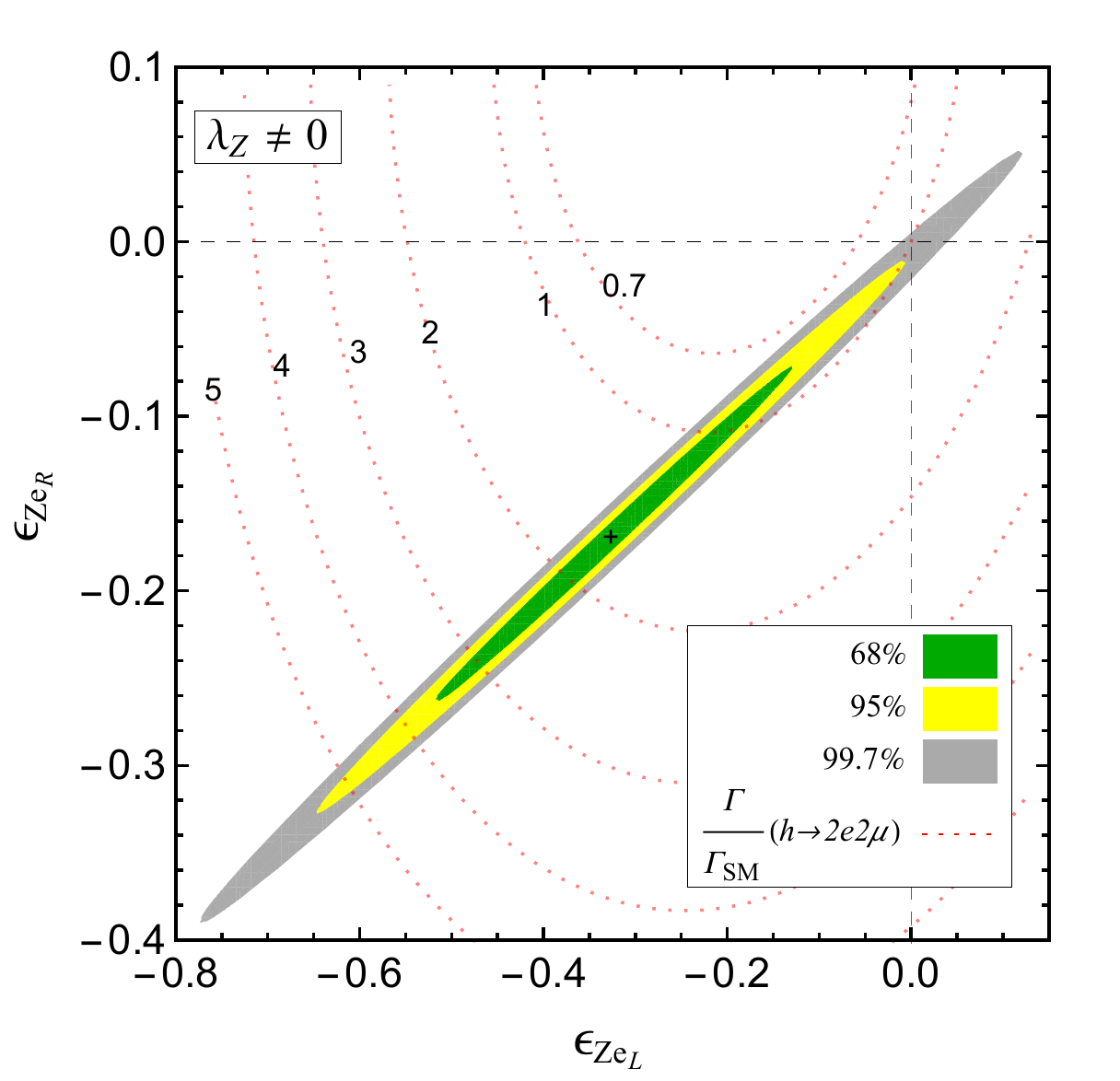} &
   \includegraphics[width=0.46\textwidth]{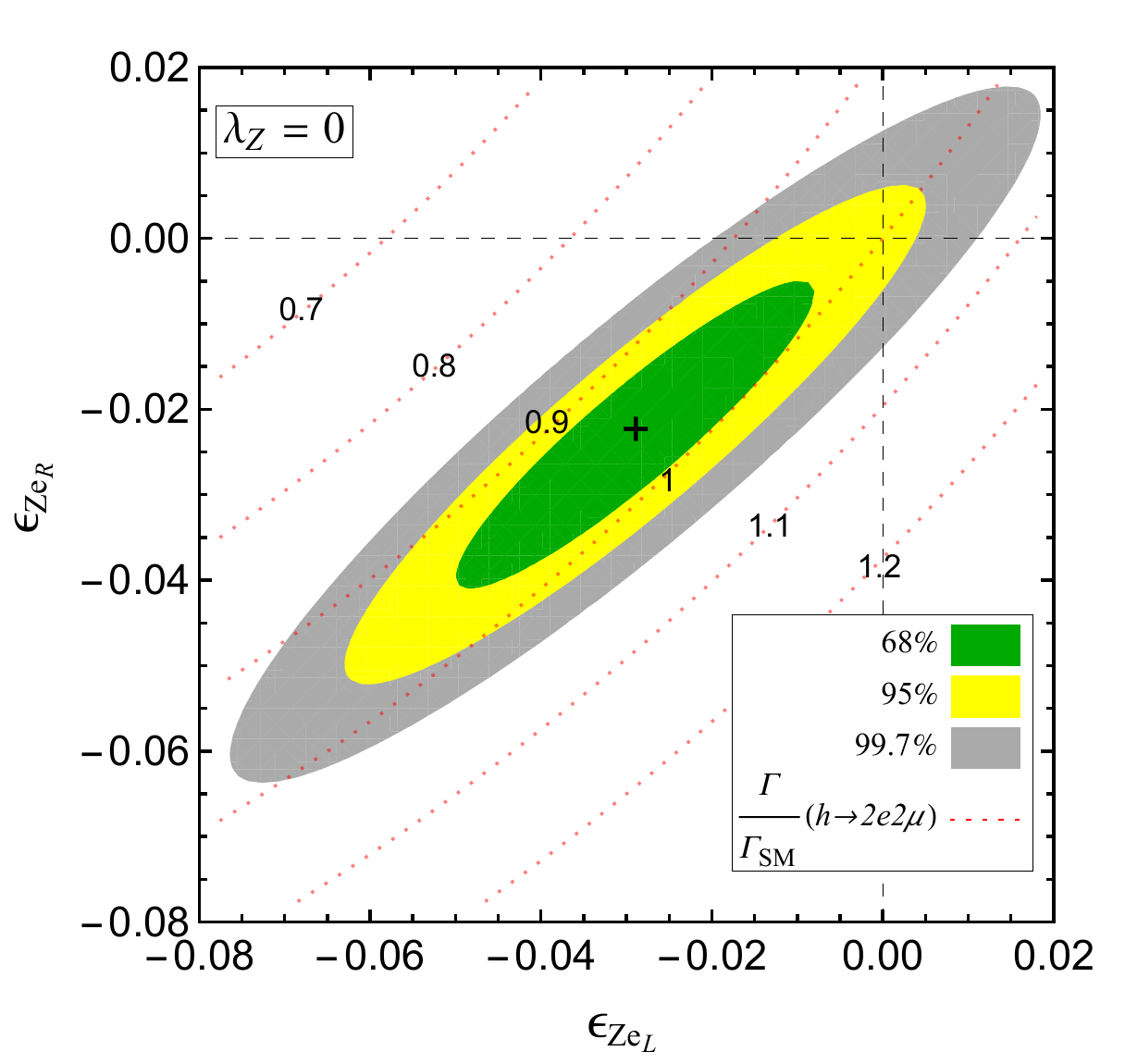}  \\
   \includegraphics[width=0.43\textwidth]{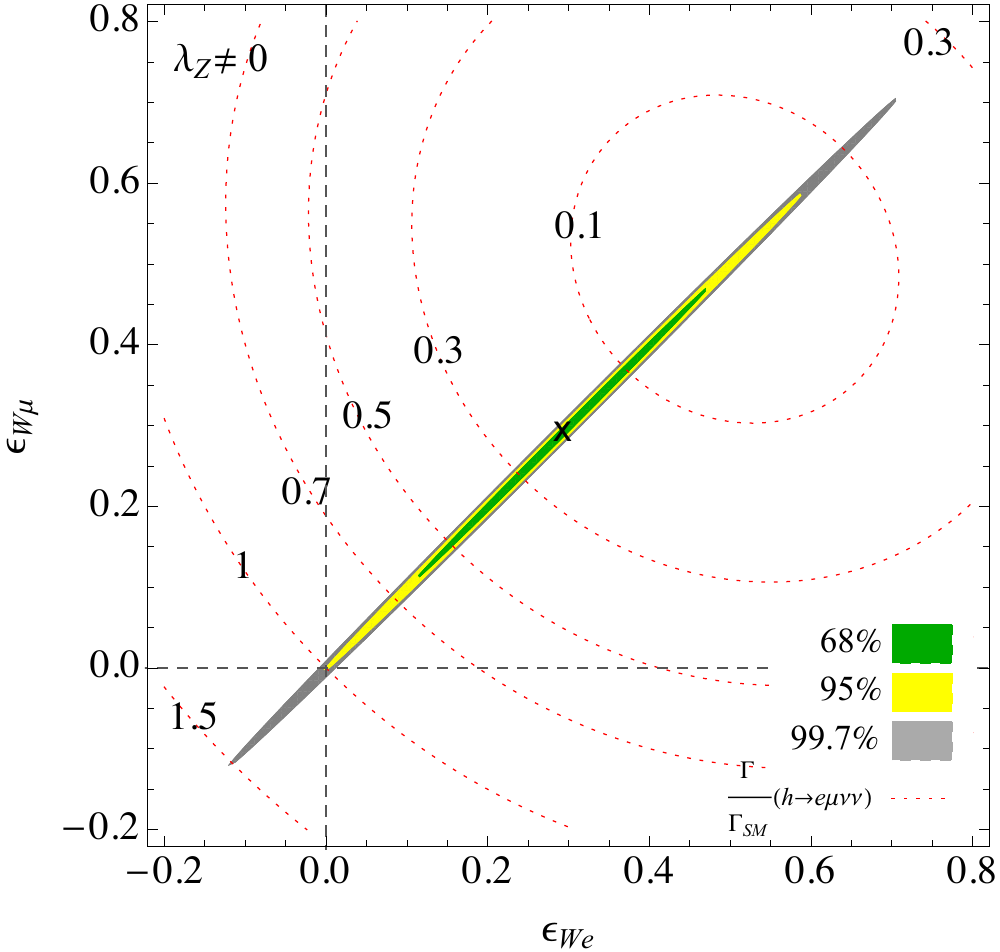} &
   \includegraphics[width=0.445\textwidth]{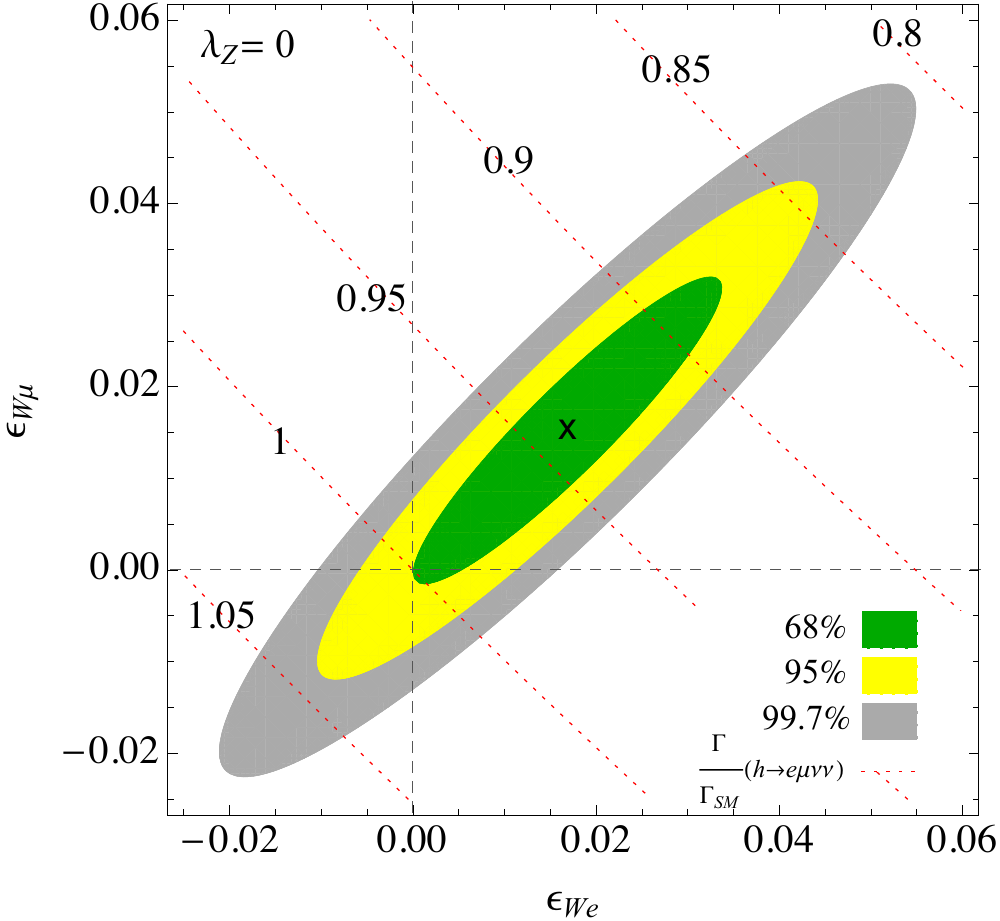}  \\
    \end{tabular}
    \caption{ \small Bounds on the contact terms $\epsilon_{Z e_L}, \epsilon_{Z e_R}$ (upper row) and $\epsilon_{W e}, \epsilon_{W \mu}$ (lower row) (at  $68\%$, $95 \%$ and $99.7 \%$ CL)
    obtained from the $W$ couplings and TGC constraints, where $\lambda_Z$ has been marginalized (left plots) or set to zero (right plots).
    The dotted contours are $\Gamma/\Gamma_{\rm SM}(h\to 2e2\mu)$ and $\Gamma/\Gamma_{\rm SM}(h\to e\mu\nu_e\nu_\mu)$ iso-lines.
    \label{fig:TGCbounds_eps}}
\end{figure*}

Given the strong bounds on $\delta g^{Z \ell}$, and the anticipated precision on the Higgs contact terms,
 we can fix the former parameters to 
their SM values in Eq.~\eqref{eq:cont_terms_EFT}  and study the allowed range of  $\epsilon_{Z f}$ and $\epsilon_{W \ell}$ as determined by the TGC 
(whose constraints are obtained in the same limit) and $W$ couplings only.

EW data also allows to bound the following custodial-symmetry-violating combination of PO~\cite{Gonzalez-Alonso:2014eva,Pomarol:2014dya},
\be\begin{split}
	&&  \kappa_{WW} - \kappa_{ZZ} + \frac{2}{g}\left( \sqrt{2} \epsilon_{W e} + 2 c_\theta \epsilon_{Z e_L} \right) \\
		&&  =  2 \delta g^{We} + 4 \delta g^{Z e_L} + 4 \delta m~,
	\label{eq:cust_symm_bound}
\end{split}\ee
where $\delta m \equiv \delta m_W / m_W$ is also constrained to be below
the per-mil level \cite{Falkowski:2014tna,Efrati:2015aaa}. 
 Substituting the contact terms from Eq.~\eqref{fig:TGCbounds_eps} one gets \cite{Pomarol:2014dya}: $\kappa_{WW} - \kappa_{ZZ} = -2 s_\theta^2 \delta g_{1,z} + 2 t_\theta^2 \delta \kappa_\gamma + 4 \delta m$.

The remaining 9 Higgs pseudo-observables  $\epsilon^{(CP)}_{ZZ,WW,Z\gamma,\gamma\gamma}$ and $\kappa_{ZZ}$
are not constrained by EW data alone. However, only five of them are independent in the linear EFT due to the following relations:
\ba
	\label{eq:epsZZdependence}	
	\delta\epsilon_{ZZ} &=& \delta\epsilon_{\gamma\gamma} +\frac{2}{t_{2\theta}} \delta\epsilon_{Z\gamma} - \frac{1}{c_\theta^2} \delta \kappa_{\gamma}~,\\
	\label{eq:epsWWdependence}	
	\delta\epsilon_{WW} &= & c_\theta^2 \delta\epsilon_{ZZ} + s_{2\theta} \delta\epsilon_{Z\gamma} + s_\theta^2 \delta\epsilon_{\gamma \gamma}~, 
\ea
and likewise for their CP counterparts (see also Refs.~\cite{Contino:2013kra,Pomarol:2014dya,HiggsBasis}). 
Here and in the following we denote by $\delta \epsilon_X$ 
 the NP contribution to the pseudo-observable $\epsilon_X$ once the one-loop SM contribution is removed: $\delta \epsilon_X = \epsilon_X - \epsilon_X^{\rm SM-1L}$.

Since no LEP bound is available 
on the CP-violating TGC coupling $\delta \tilde{\kappa}_\gamma$, at present $\epsilon^{CP}_{ZZ}$ is an independent variable. However, in the future 
 significant constraints on  $\delta \tilde{\kappa}_\gamma$ could be obtained from LHC data~\cite{Dawson:2013owa}.
 All in all, we are left with 3 CP-con\-serving couplings, $\kappa_{ZZ}$ and $\epsilon_{\gamma\gamma,Z\gamma}$, and 3 CP-violating ones, 
 $\epsilon^{CP}_{\gamma\gamma,Z\gamma,ZZ}$.

In principle, the measurements of the partial decay widths $\Gamma(h \rightarrow \gamma\gamma, Z\gamma)$   allow to set strong bounds on $\epsilon^{(CP)}_{\gamma\gamma,Z\gamma}$ that, when combined with the TGC bounds,  imply strong limits on $\epsilon_{ZZ,WW}$ through 
Eqs.~\eqref{eq:epsZZdependence} and \eqref{eq:epsWWdependence}. 
In practice, the extraction of such bounds is not straightforward since, at present, only the measurements of the so-called signal strengths (or $\sigma \times BR$ normalized to SM) are available. The latter include also possible non-standard effects in the Higgs production and in the total decay width (e.g. via $\kappa_{ZZ}\neq 1$). We benefit from various global fits available in the literature~\cite{Pomarol:2013zra,atlas-combination,Khachatryan:2014kca,Khachatryan:2014jba}, which imply per-mil level limits on $\epsilon^{(CP)}_{\gamma\gamma}$ and per-cent level limits on $\epsilon^{(CP)}_{Z\gamma}$.
In particular, in the following we use the values~\cite{atlas-combination}
\be
	\kappa_{\gamma\gamma} = 0.90 \pm 0.15, \qquad  |\kappa_{Z\gamma}| < 3.18 ~ (95\%~{\rm CL})~,
	\label{eq:gg_Zg_ATLASbounds}
\ee
where $\kappa_{\gamma\gamma,Z\gamma} \equiv \epsilon_{\gamma\gamma,Z\gamma} / \epsilon_{\gamma\gamma,Z\gamma}^{\rm SM-1L}$, with 
$\epsilon_{\gamma\gamma}^{\rm SM-1L} \approx 3.8 \times 10^{-3}$, and $\epsilon_{Z\gamma}^{\rm SM-1L} \approx 6.7 \times 10^{-3}$. As discussed above, the constraints on $\epsilon_{\gamma\gamma,Z\gamma}^{CP}$ are equivalent to those shown above for their CP-conserving counterparts $\epsilon_{\gamma\gamma,Z\gamma}$, whereas no bound is available for $\kappa_{ZZ}$ and $\epsilon_{ZZ}^{CP}$ (before analyzing $h\to 4\ell$ data).

 Combining the bounds on the TGC with those on the $W$ couplings to leptons,
we find the following constraints on the Higgs contact terms 
\ba
\label{eq:boundsPO} 
&\left ( \begin{array}{c} 
\epsilon_{Z e_L}   \\
\epsilon_{Z e_R}    \\ 
\epsilon_{W e}   \\
\epsilon_{W \mu}
\end{array} \right )  = 
\left ( \begin{array}{c} 
-0.32(13) \\ 
-0.17(6) \\ 
0.29(12) \\
0.29(12)
\end{array} \right )_{\!\!\!\!\lambda_Z\neq0}
\!\!\!\!\!\!\!\!\!\!= 
\left ( \begin{array}{c} 
-0.029(14) \\ 
-0.023(12) \\ 
0.017(11) \\
0.015(11)
\end{array} \right )_{\!\!\!\!\lambda_Z=0}
\!\!\!,
\label{eq:boundsPOlamZ}
\ea
with the following correlation matrix
\ba
\rho &=& 
\left ( \begin{array}{cccc} 
1 &  0.9961 &  -0.9993 &  -0.9993 \\ 
\cdot   & 1  & -0.9929 &  -0.9929 \\ 
\cdot   & \cdot &  1 &  0.9994 \\ 
\cdot   & \cdot & \cdot & 1
\end{array} \right )_{\!\!\!\!\lambda_Z\neq0} 
\nonumber\\
&=&
 \left ( \begin{array}{cccccc} 
1 &  0.92 &  -0.93 &  -0.93 \\ 
\cdot   & 1  & -0.75 &  -0.76 \\ 
\cdot   & \cdot &  1 &  0.93 \\ 
\cdot   & \cdot & \cdot & 1 
\end{array} \right )_{\!\!\!\!\lambda_Z=0} ~.
\label{eq:boundsPOlamZ2}
\ea
Up to per-mil corrections due to deviations in the $Z$ couplings to leptons and in $\delta m$, the following relations among Higgs PO are satisfied:
\ba
\label{eq:epsZZ}
	\delta\epsilon_{ZZ} &=& \delta\epsilon_{\gamma\gamma} + \frac{2}{t_{2\theta}} \delta\epsilon_{Z\gamma} - \frac{v}{c_\theta^2 m_Z} \left( \epsilon_{Z e_L} - \frac{1}{2 s_\theta^2} \epsilon_{Z e_R} \right)~ \nonumber \\
	\delta\epsilon_{WW} &=& \delta\epsilon_{\gamma\gamma} + \frac{1}{t_\theta} \delta\epsilon_{Z\gamma} - \frac{v}{m_Z} \left( \epsilon_{Z e_L} - \frac{1}{2 s_\theta^2} \epsilon_{Z e_R} \right)~\\
	\kappa_{WW} - \kappa_{ZZ} &=& -\frac{v}{m_Z} \epsilon_{Z e_R}~, \nonumber
\ea
which can be used to derive constraints on these PO given the bounds in Eqs.~(\ref{eq:gg_Zg_ATLASbounds},\ref{eq:boundsPOlamZ},\ref{eq:boundsPOlamZ2}).

In the first row of Fig.~\ref{fig:TGCbounds_eps} we present the bounds on the $\epsilon_{Z e_L}$ and $\epsilon_{Z e_R}$ pseudo-observables (relevant for $h \rightarrow 2e2\mu$, $4e$, $4\mu$ decays) both in the general case  ($\lambda_Z\not=0$,  marginalised) and for $\lambda_Z=0$.
It is interesting to notice that, even in the general case, only the direction
\be
\epsilon_{Z e_R} \approx 0.48 \times \epsilon_{Z e_L}
\label{eq:ZLZRapp}
\ee 
is loosely bounded, and that sizable positive values of the contact terms are excluded. The particular flat direction in the contact terms can be understood analytically by the fact that 
$ \delta \kappa_\gamma$ is much more constrained than $ \delta g_{1,z}$. As a result, we can also set $\delta \kappa_\gamma \approx 0$
in Eq.~\eqref{eq:cont_terms_EFT}, which implies  $\epsilon_{Z e_R} \approx  2 s_\theta^2 \epsilon_{Z e_L} \approx 0.46 \times \epsilon_{Z e_L}$
(up to a $\sim 10\%$ accuracy). 

In the second row of Fig.~\ref{fig:TGCbounds_eps} we show the constraints on $\epsilon_{W e}$ and $\epsilon_{W \mu}$ (relevant for $h \rightarrow e\mu \nu_e \nu_\mu, 2e2\nu_e, 2\mu 2\nu_\mu$ decays) for both hypotheses on $\lambda_Z$. In particular we notice that in the general $\lambda_Z \neq 0$ case, since the bound on $\delta g_{1,z}$ is much worse than those on $\kappa_\gamma$ and $\delta g^{We}, \delta g^{W \mu}$, all four contact terms in Eq.~\eqref{eq:boundsPOlamZ} are highly correlated since -- effectively -- they all depend on the single variable $\delta g_{1,z}$. This implies two further relations: 
\ba
&& \epsilon_{W e} \approx \epsilon_{W \mu} \\
&& \epsilon_{W e} \approx \frac{v^2}{\sqrt{2} m_W m_Z} (\epsilon_{Z e_R} - \epsilon_{Z e_L}).
\ea
in addition to the one in Eq.~\eqref{eq:ZLZRapp}.

In the rest of this work we study the implications of these constraints for $h\to4\ell~ (\ell=e,\mu)$ decay rates and dilepton invariant-mass distributions. 
More specifically, we will propagate the errors shown above and analyze the allowed room for non-standard effects. 
We include quadratic terms in all following calculations, which in general should represent subleading corrections in the EFT expansion. However, this is not always the case in the general scenario $\lambda_Z \neq 0$, since values as large as 0.4 are allowed for some of the pseudo-observables.
The subsequent predictions for $h\to4\ell$  observables should be taken with care in this case, and interpreted as the room for New Physics taking into account that very little is known on certain pseudo-observables. 

\begin{figure*}[t]
\ba
\qquad\qquad\qquad
X &=&
	\left(
\begin{array}{ccccccccc}
 1.0 & 3.0 & -2.4 & 3.0 & -2.4 & 0.55 & -1.1 & 0.021	& 0 \\
0 & 6.2 & 0 & 1.7 & -1.4 & 0.80 & -13 & 0.36 & 0  \\
 0 & 0 & 6.2 & -1.4 & 1.1 & -0.64 & -13 & 0.36 & 0 \\
 0 & 0 & 0 & 6.2 & 0 & 0.80 & -13 & 0.36 & 0\\
 0 & 0 & 0 & 0 & 6.2 & -0.64 & -13 & 0.36 & 0 \\
 0 & 0 & 0 & 0 & 0 & 0.099 & -0.39 & 0.0094 & 0 \\
 0 & 0 & 0 & 0 & 0 & 0 & 52 & -1.9 & 0 \\
 0 & 0 & 0 & 0 & 0 & 0 & 0 & 31 & 0 \\
 0 & 0 & 0 & 0 & 0 & 0 & 0 & 0 & X_{3\times3}^{CP} \\
\end{array}
\right)_{\!\!\!\!\!2e2\mu}~
\!\!\!\!\!\!\!\!\!\!\!\!\!=
	\left(
\begin{array}{ccccccccc}
 1.0 & 6.3 & -5.0 & 0 & 0 & 0.52 & -0.89 & 0.50 & 0.0 \\
 0 & 16 & -2.8 & 0 & 0 & 1.6 & -25 & 2.9 & 0.0 \\
 0 & 0 & 15 & 0 & 0 & -1.3 & -25 & -1.1 & 0.0 \\
 0 & 0 & 0 & 0 & 0 & 0 & 0 & 0 & 0\\
 0 & 0 & 0 & 0 & 0 & 0 & 0 & 0 & 0\\
 0 & 0 & 0 & 0 & 0 & 0.085 & -0.27 & -0.07 & 0.0 \\
 0 & 0 & 0 & 0 & 0 & 0 & 44 & -1.6 & 0.0 \\
 0 & 0 & 0 & 0 & 0 & 0 & 0 & 26 & 0.0 \\
 0 & 0 & 0 & 0 & 0 & 0 & 0 & 0 & X_{3\times3}^{CP} \\
\end{array}
\right)_{\!\!\!\!\!4e}
\label{eq:total_rate_12GeV}
\\
\qquad\qquad X_{3\times3}^{CP} &=&
	\left(
\begin{array}{ccc}
0.04 & -0.20 & 0.007 \\
0 & 34 & -1.6 \\
0 & 0 & 29 \\
\end{array}
\right)_{\!\!\!\!\!2e2\mu}~
\!\!\!\!\!\!\!\!=
	\left(
\begin{array}{ccc}
0.033 & -0.14 & -0.18 \\
0 & 27 & -1.1 \\
0 & 0 & 23 \\
\end{array}
\right)_{\!\!\!\!\!4e}
\label{eq:total_rate_12GeVbis}
\ea
\end{figure*}

\begin{figure*}[t]
  \centering
  \begin{tabular}{cc}
   \includegraphics[width=0.47\textwidth]{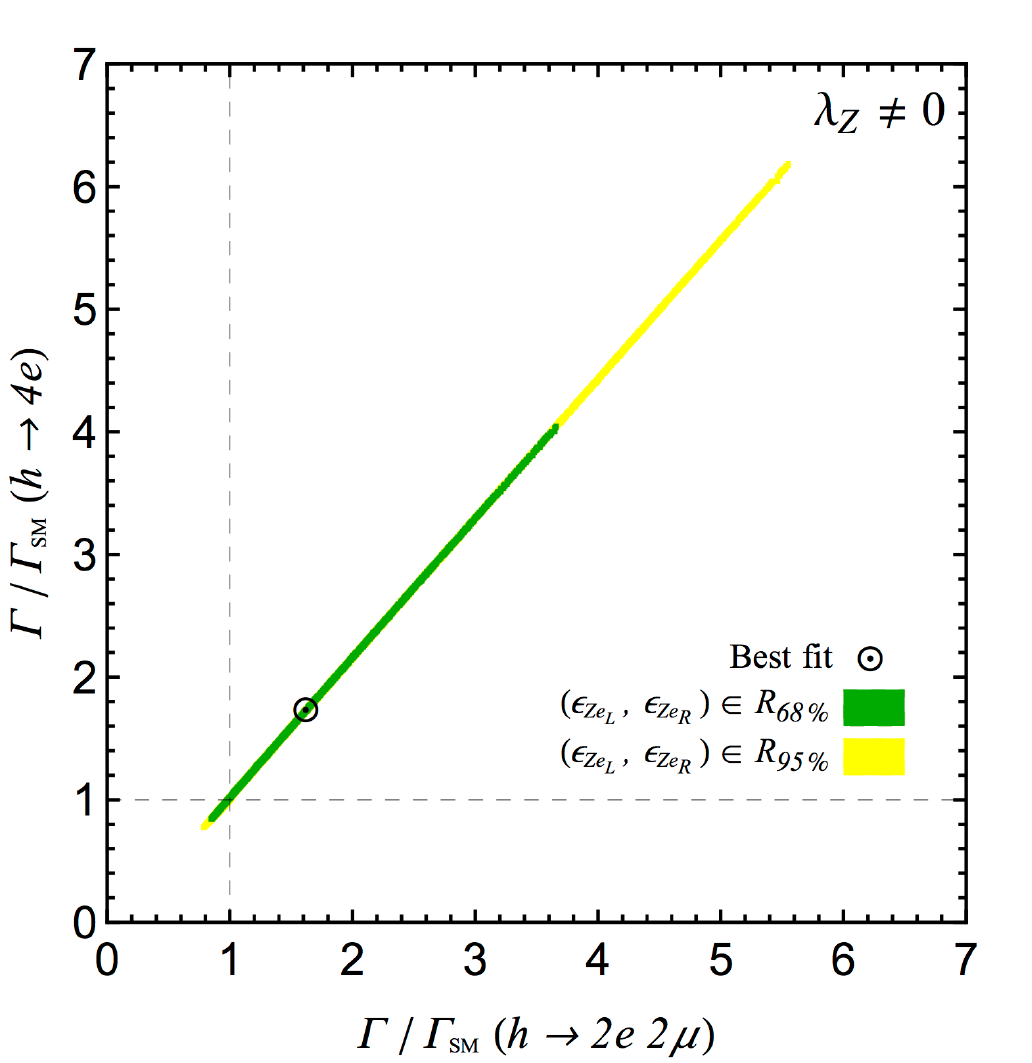} &
   \includegraphics[width=0.50 \textwidth]{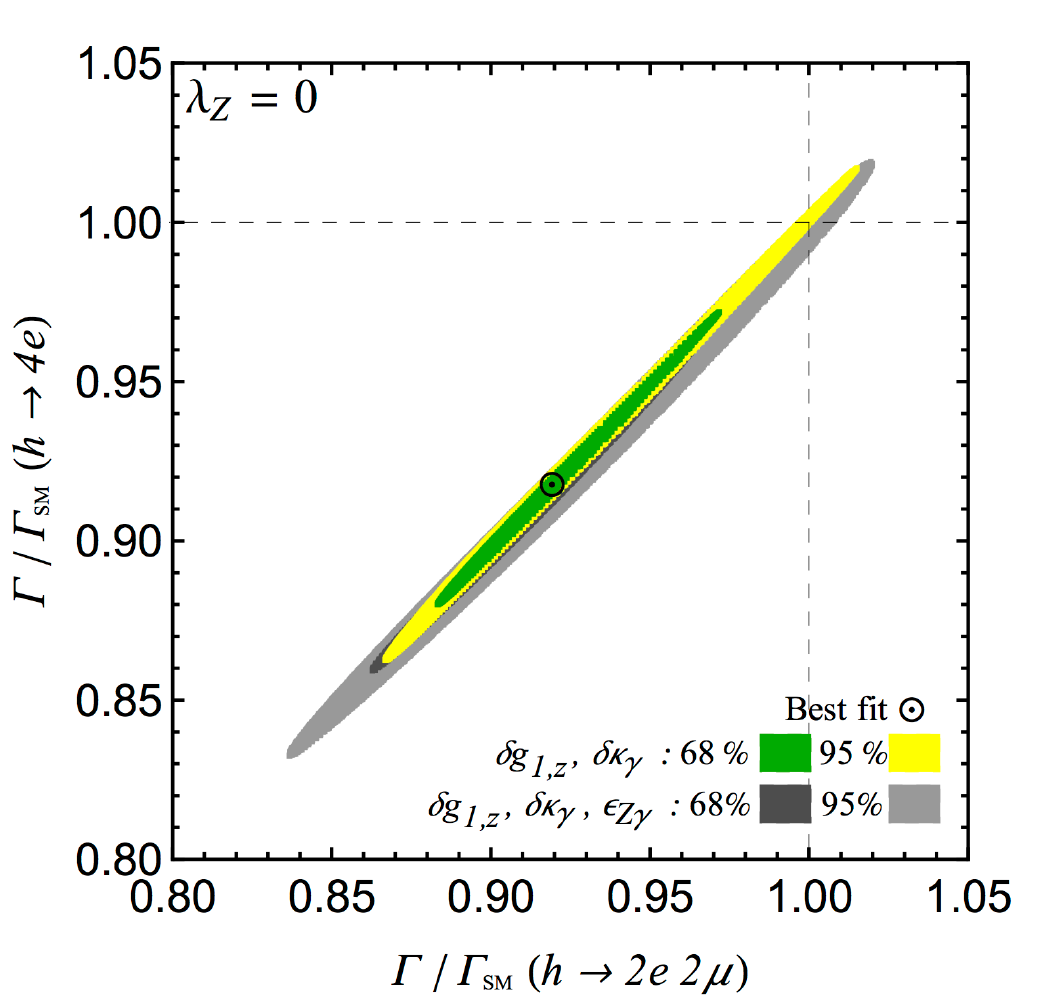}  \\
    \end{tabular}
    \caption{ \small Predictions for $h\to4e$ versus $h\to2e2\mu$ decay rates implied by TGC constraints.
    Left: $\lambda_Z\not=0$ case. Right: $\lambda_Z=0$ case. 
   The results obtained varying $\epsilon_{Z e_L}$ and 
    $\epsilon_{Z e_R}$ only  (via $\delta g_{1,z}$ and $\delta \kappa_\gamma$), according to Fig.~\ref{eq:cont_terms_EFT}
    are  shown in green ($68\%$ CL) and yellow ($95\%$ CL).
   The additional impact of varying $\epsilon_{Z\gamma}$ within its current limits is shown in
   dark ($68\%$ CL) and light gray ($95\%$ CL) in the right plot.  In both plots we have set  $\kappa_{ZZ}=1$.  \label{fig-rates-correlations}}
\end{figure*}

\section{$h\to 4\ell$ phenomenology under EW constraints}
\subsection{Partial decay rates}

We compute the modification of the $h\to4\ell$ integrated decay rates due to non-standard contributions to the pseudo-observables.
In order to regulate the photon pole and simultaneously resemble the realistic present analysis~\cite{Khachatryan:2014kca,atlas-combination}, we employ a minimum invariant-mass cut on the opposite sign same flavor lepton pairs of $m_{\ell\ell}^{\rm min}={\rm 12~GeV}$. This way we determine 
\be
\frac{\Gamma_{4\ell}}{\Gamma_{4\ell}^{\rm SM}} = \sum_{i,j} X^{4\ell}_{ij} \kappa_i \kappa_j~,
\label{eq:total_rate_h4l}
\ee
where
\be
\kappa \equiv \left\{ \kappa_{ZZ},\epsilon_{Z e_{L}},\epsilon_{Z e_{R}},\epsilon_{Z \mu_{L}},\epsilon_{Z \mu_{R}},\epsilon_{ZZ},\epsilon_{Z \gamma},\epsilon_{\gamma \gamma},\cpeps_{ZZ},\cpeps_{Z\gamma},\cpeps_{\gamma \gamma} \right\} \nonumber
\ee
and $X^{2e2\mu}, X^{4e}$ are given in Eqs.~\eqref{eq:total_rate_12GeV}--\eqref{eq:total_rate_12GeVbis} ($X^{4\mu}$ is trivially obtained from $X^{4e}$). 
The measurements of the integrated decay rates constrain only these particular PO combinations. 

Some comments on these expressions are in order.
First, it is easy to see that the contributions from the CP-violating terms $\epsilon_{ZZ,Z\gamma,\gamma\gamma}^{CP}$ are negligible once the constraints from $\Gamma(h \rightarrow \gamma\gamma, Z\gamma)$ are taken into account.
We stress that this conclusion holds even for 
$m_{\ell\ell}^{\rm min}$
as low as ${\rm 1~GeV}$. Indeed, despite the lack of bounds on $\epsilon_{ZZ}^{CP}$, its contribution to the total
rate is below 4\% even for $O(1)$ values. Thus,  $\Gamma(h\to4\ell)$ can be expressed in terms of $\kappa_{ZZ}$ and the 5 pseudo-observables 
$\epsilon_{Ze_R,Ze_L,ZZ,Z\gamma,\gamma\gamma}$ bounded by Eqs.~\eqref{eq:gg_Zg_ATLASbounds}, \eqref{eq:boundsPO} and \eqref{eq:epsZZ}.

The global fit of Ref.~\cite{atlas-combination}, that allows approximately for 30\% non-standard contributions in $\Gamma(h\to4\ell)$,
can in principle be used to obtain a bound on  $\kappa_{ZZ}$ via Eq.~\eqref{eq:total_rate_h4l}. 
However,  the error in the contact terms gets significantly enhanced when propagated to the total rate, which makes difficult to set a meaningful bound on 
$\kappa_{ZZ}$ at this point. Moreover, the fit of Ref.~\cite{atlas-combination} assumes SM-like differential spectra in $h\to4\ell$, that is not necessarily a safe assumption in the presence of sizable contact terms. We will come back to  the combined bounds 
on $\kappa_{ZZ}$ and the contact terms from the partial rates at the end of this section, 
after addressing the possible non-standard effects on the dilepton invariant-mass spectra.

\begin{figure*}[p]
  \centering
  \begin{tabular}{cc}
   \hspace{-0.5cm} \includegraphics[width=0.5\textwidth]{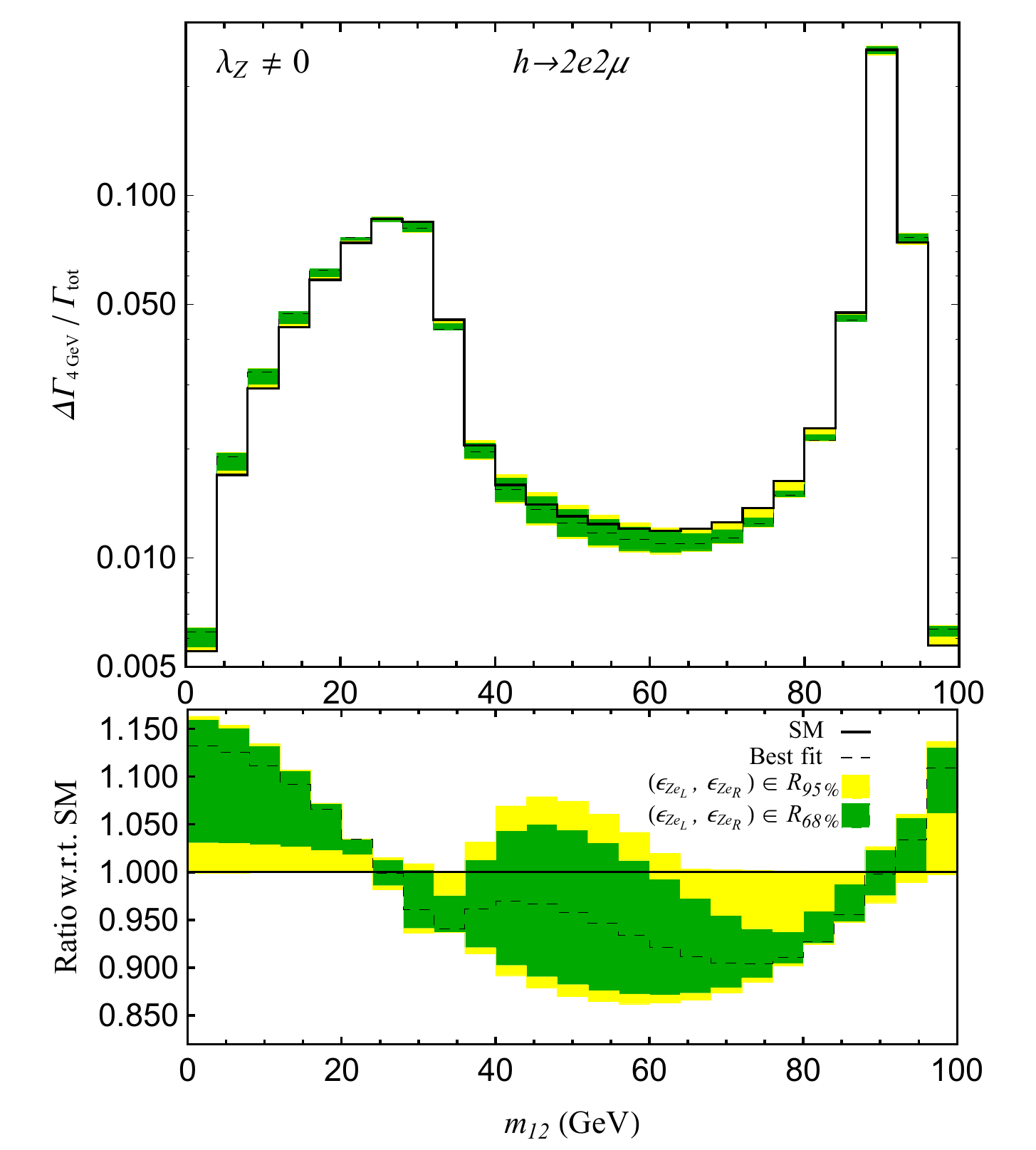} &
    \includegraphics[width=0.525\textwidth]{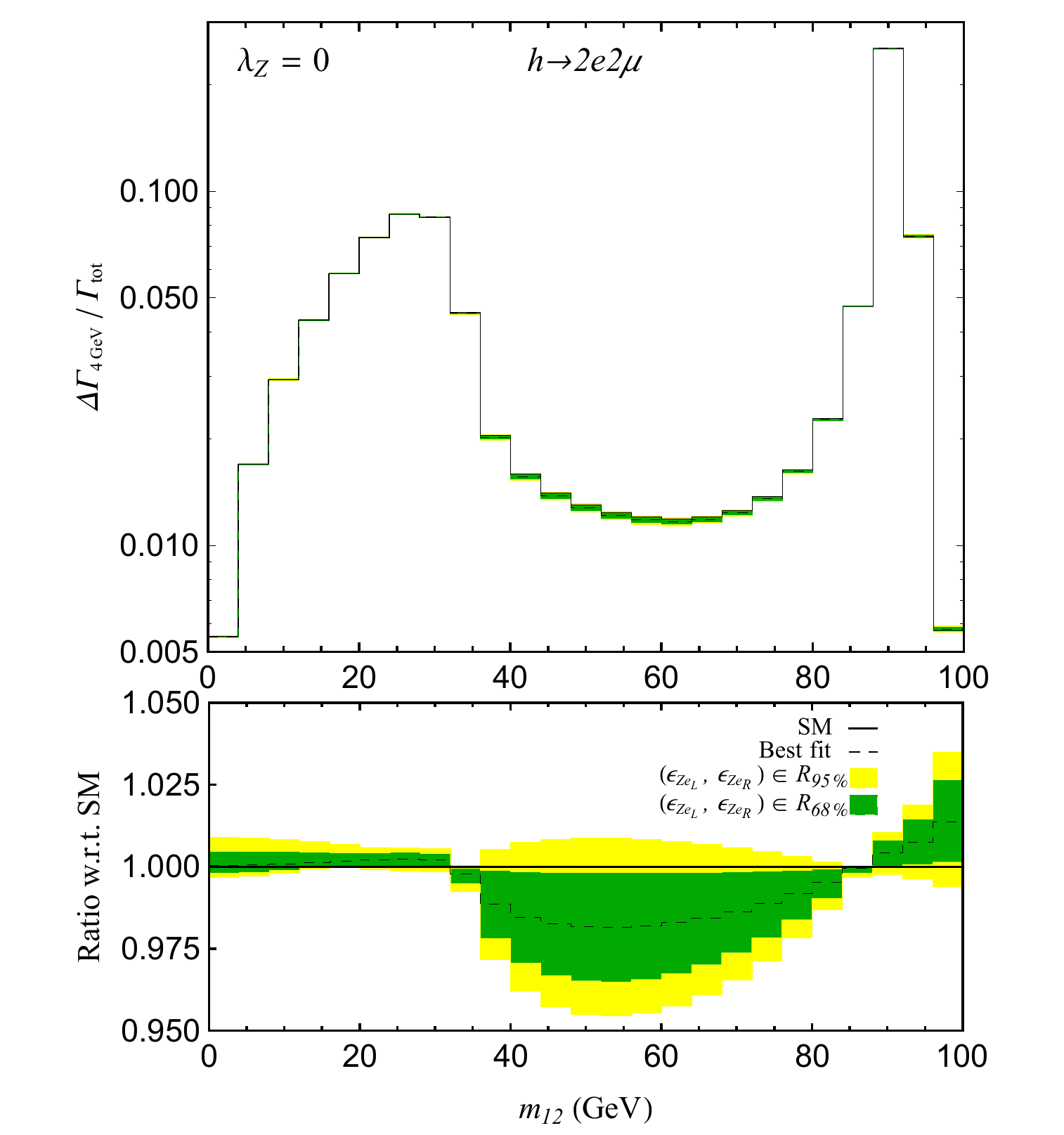} \\
            (a) & (b)\vspace{0.5cm}\\ 
    \hspace{-0.5cm} \includegraphics[width=0.5\textwidth]{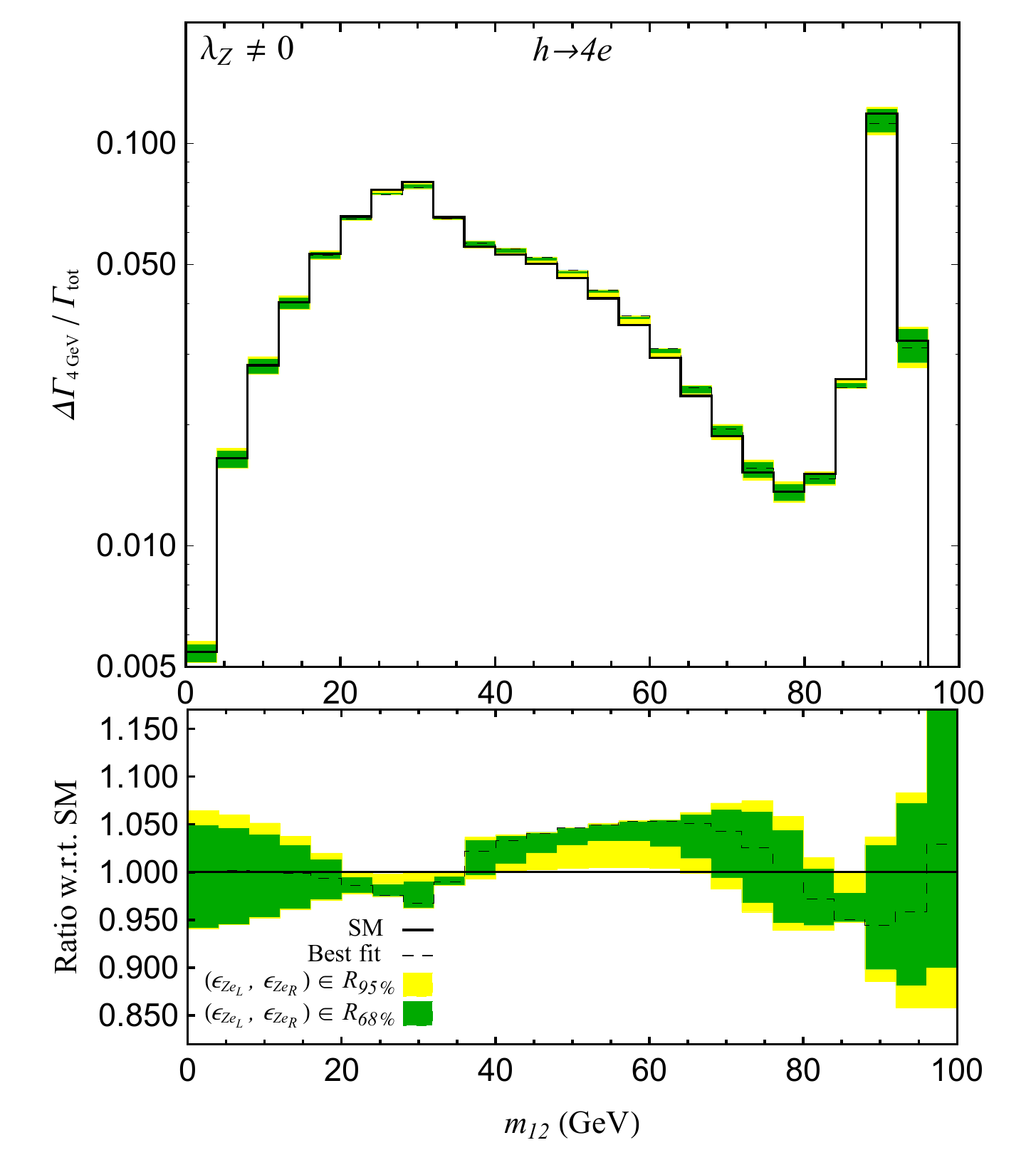} &
    \includegraphics[width=0.5\textwidth]{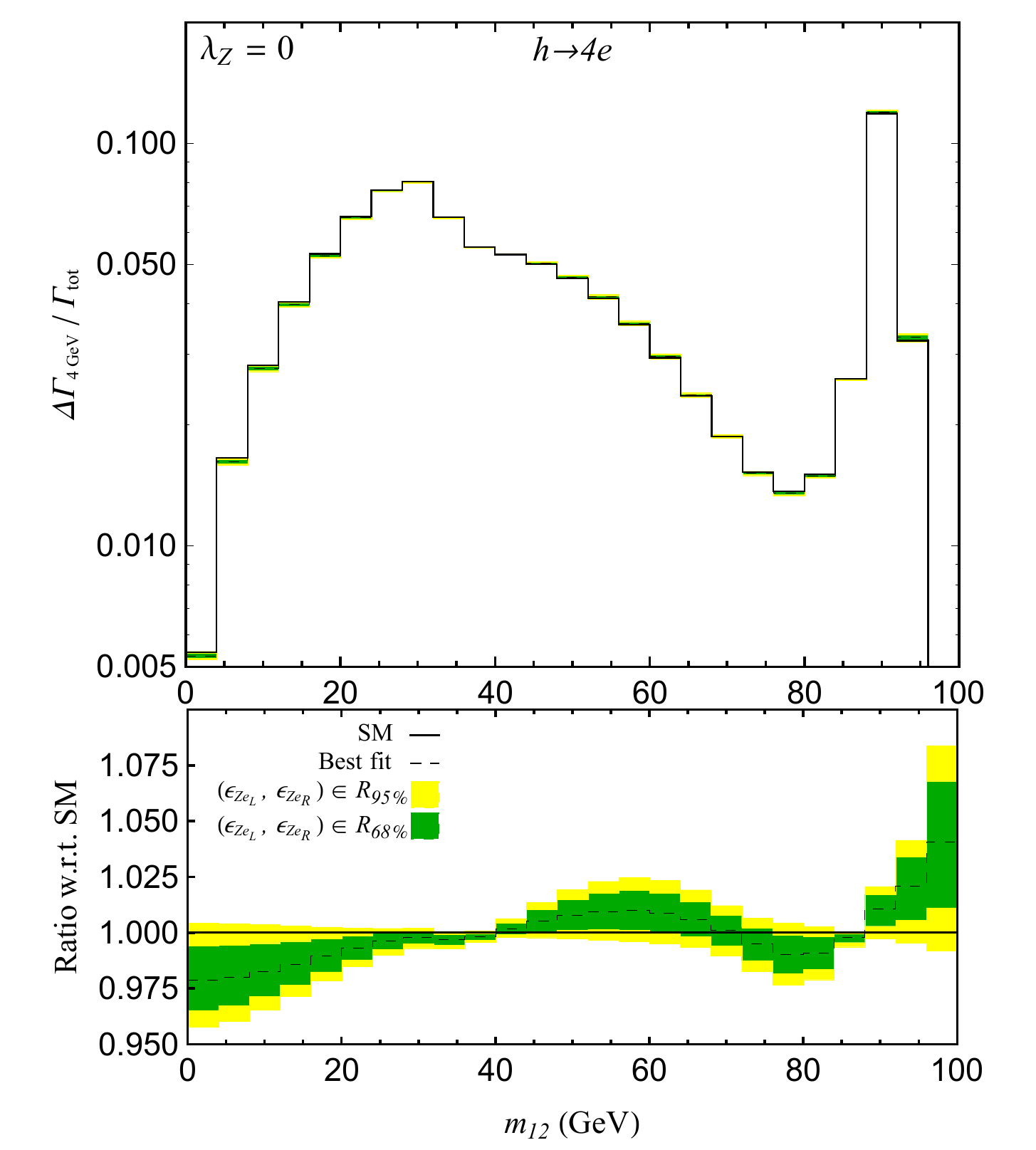} \\
            (c) & (d)\\
    \end{tabular}    \caption{\label{fig:Diff_distr_cont_terms} \small Electron pair invariant mass spectrum with a 4 GeV binning for 
    $h\to 2e2\mu$ (top row) and $h \to 4e$ (bottom row) decay obtained by varying $\epsilon_{Z e_L}$ and $\epsilon_{Z e_R}$ within the $68\%$ (green) and $95\%$ (yellow) CLs bound from TGC (Fig.~\ref{fig:TGCbounds_eps}) with $\lambda_Z$ generic in Fig.~(a,c), and  
    $\lambda_Z=0$ in Fig.~(b,d).  In the $h\to4e$ channel we pair randomly two opposite-sign leptons. }
\end{figure*}

The dependence of the partial widths, $\Gamma(h\to4\ell)$, from all the PO but for $\kappa_{ZZ}$ is illustrated in Fig.~\ref{fig-rates-correlations}. 
In the left plot we consider the general TGC case (no assumptions on $\lambda_Z$). As can be seen, $\mathcal{O}(1)$ variations on the rates are allowed because of the weak bounds on the contact terms (see Fig.~\ref{fig:TGCbounds_eps} left).
However, a tight correlation between $h\to 2e2\mu$ and $h\to 4e (4\mu)$ rates holds because of the flavor universality implied by LEP data under the hypothesis that $h(125)$ belongs to a pure $\SU(2)_L$ doublet. 
In the right plot we consider the $\lambda_Z=0$ case: in this limit the overall modifications are much reduced but still visible, while the correlation between $h\to 2e2\mu$ and $h\to 4e (4\mu)$ remains very strong, with possible deviations below any future realistic resolution. 

The different effect of the photon pole in the two channels, discussed in Ref.~\cite{Chen:2015iha},
manifests itself in Eq.~\eqref{eq:total_rate_12GeV} as a $\sim 25$ times larger interference term 
of $\epsilon_{\gamma\gamma}$ with $\kappa_{ZZ}$ in $h \rightarrow 4e(4\mu)$ vs.~$h \rightarrow 2e2\mu$.
However, we stress that this is a tiny effect on the partial rates (below $1\%$ with present cuts)
once the LHC bound on $\epsilon_{\gamma\gamma}$ is taken into account. This is why this effect is not visible in Fig.~\ref{fig-rates-correlations}. The smallness of this effect also 
implies that improving the bounds on $\epsilon_{\gamma\gamma}$ from $h\to   4e (4\mu)$ is extremely challenging, 
especially in the general case where the SM deviations from all the PO are considered at the same time.

The strict correlation between $h\to 2e2\mu$ and $h\to 4e (4\mu)$ rates represents a firm prediction of the linear EFT frameworks that is  
worth to test with future data: any violation of the correlation would not only imply the existence of NP, but would also imply that 
i)~NP does not respect lepton universality, ii)~the Higgs particle has a non-$\SU(2)_L$ component.\footnote{We stress that 
these two conditions are not sufficient to ensure large deviations from universality in $h\to 4\ell$ decays, 
but are necessary conditions to observe it.}

\subsection{Single dilepton invariant-mass spectra}

In addition to the partial widths, the rich kinematics of the $h\to4\ell$ processes offers additional handles to probe the relevant pseudo-observables. Since the contact terms $\epsilon_{Ze_L,Ze_R}$ have the same Lorentz structure as the SM term, angular distributions are not modified and the only effect is on the differential distributions in the dilepton invariant masses. On the other hand, the other pseudo-observables, $\epsilon_{ZZ,Z\gamma,\gamma\gamma}^{(CP)}$, 
modify also angular distributions and thus a complete study of the full kinematics of the events is necessary in order to extract them as efficiently as possible
(see in particular Refs.~\cite{Buchalla:2013mpa,Gainer:2014hha,Chen:2014gka,Beneke:2014sba,Chen:2015iha} for a recent discussion). 
In this work  we focus only on the invariant-mass distributions, both because the effect of the contact terms in  $h\to4\ell$ is the one less studied in the previous literature and because, as shown above, these PO are the less constrained at the moment (at least in the general TGC case).

\begin{figure}[t]
  \centering
    \hspace{-0.5cm}
    \includegraphics[width=0.50\textwidth]{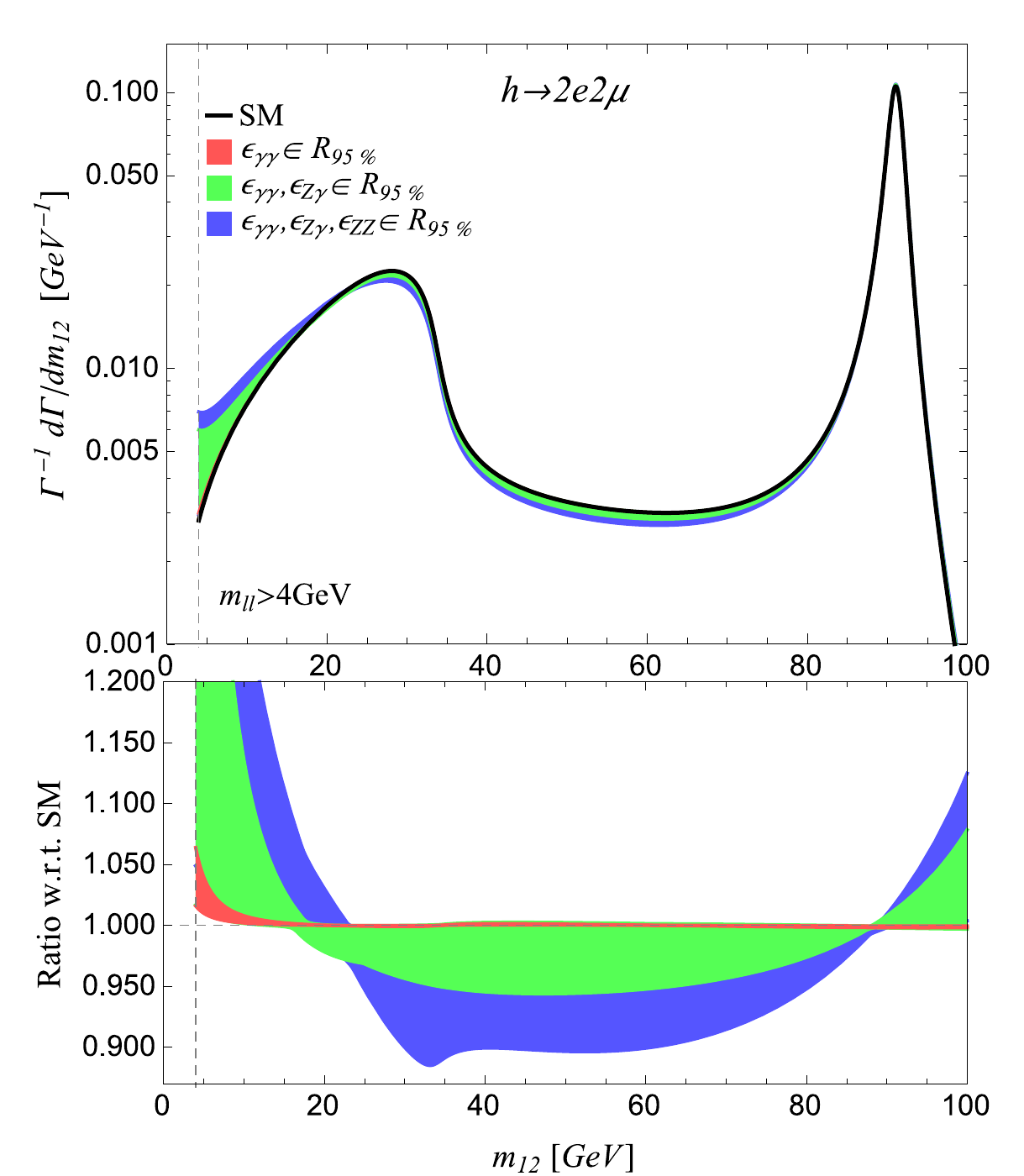}
    \caption{\label{fig:Diff_distr_eps_ggZgZZ} \small Single differential distributions in the electron pair invariant mass for $h\to 2e2\mu$ decay obtained by varying $\epsilon_{\gamma\gamma}$, $\epsilon_{Z\gamma}$ and $\epsilon_{ZZ}$ inside the $95\%$ CL bounds obtained from Eqs.~(\ref{eq:gg_Zg_ATLASbounds}) and (\ref{eq:epsZZdependence}) and setting $\epsilon_{Z\ell}=0$, $\kappa_{ZZ} = 1$.
        A lower cut on both lepton pair's invariant masses of 4 GeV is applied. In the upper plot the differential rate is normalized to the total rate while in the lower one we take the ratio of this quantity to the one obtained in the SM at the tree level.}
\end{figure}

Since the effects on the partial widths have already been discussed,  here we focus on the shapes, i.e.~the normalized differential distributions.
To this purpose, we have sampled sets of PO inside the $68\%$ and $95\%$ CL bounds, keeping into account their correlations. 
For each set we have determined the normalized dilepton invariant-mass spectrum and its ratio to the one obtained in the SM at tree level, 
and  we have finally built the envelopes of 
such spectra.

\begin{figure*}[t]
  \centering
  \begin{tabular}{cc}
   \hspace{-0.5cm} \includegraphics[width=0.50\textwidth]{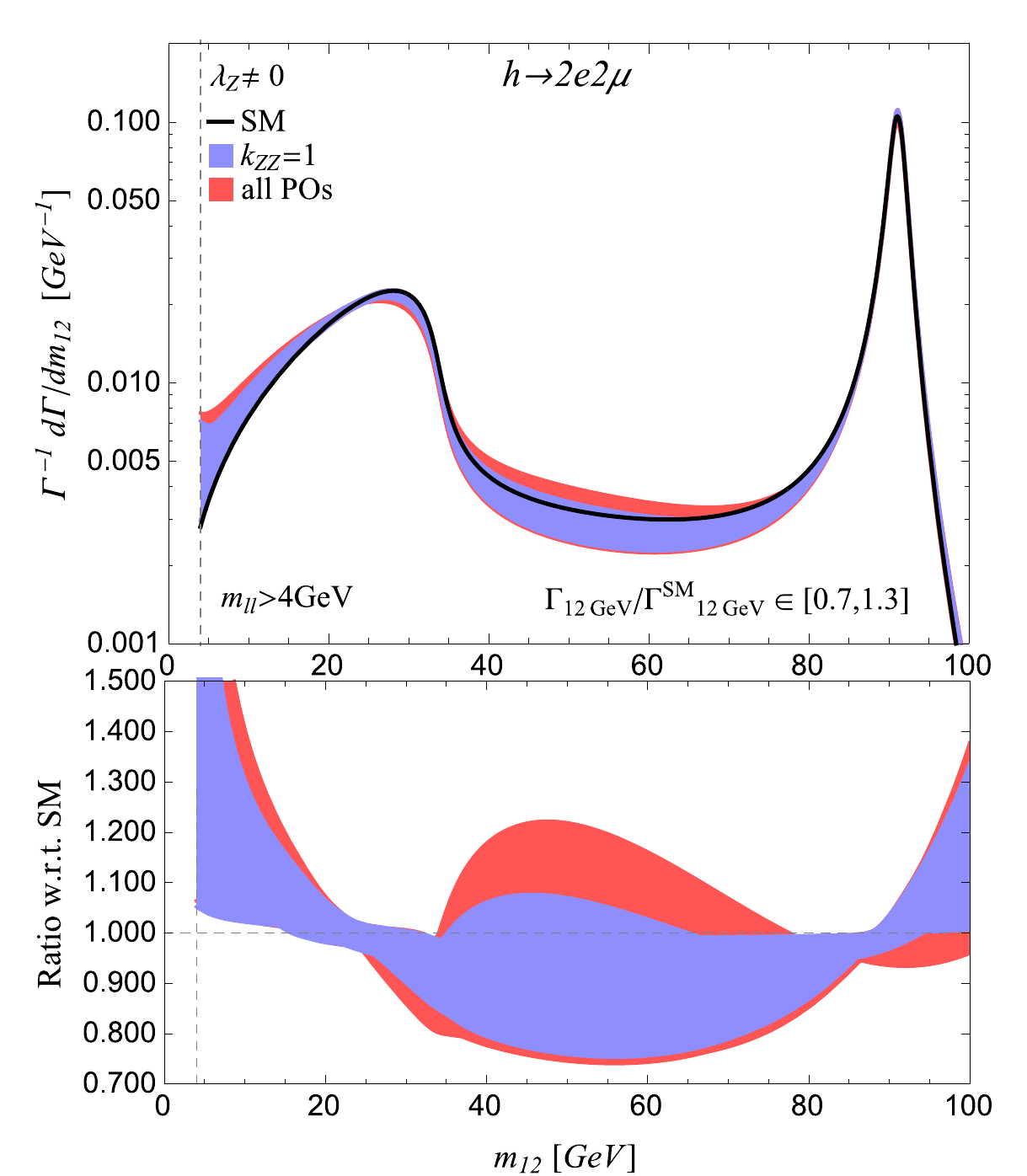} &
    \includegraphics[width=0.50\textwidth]{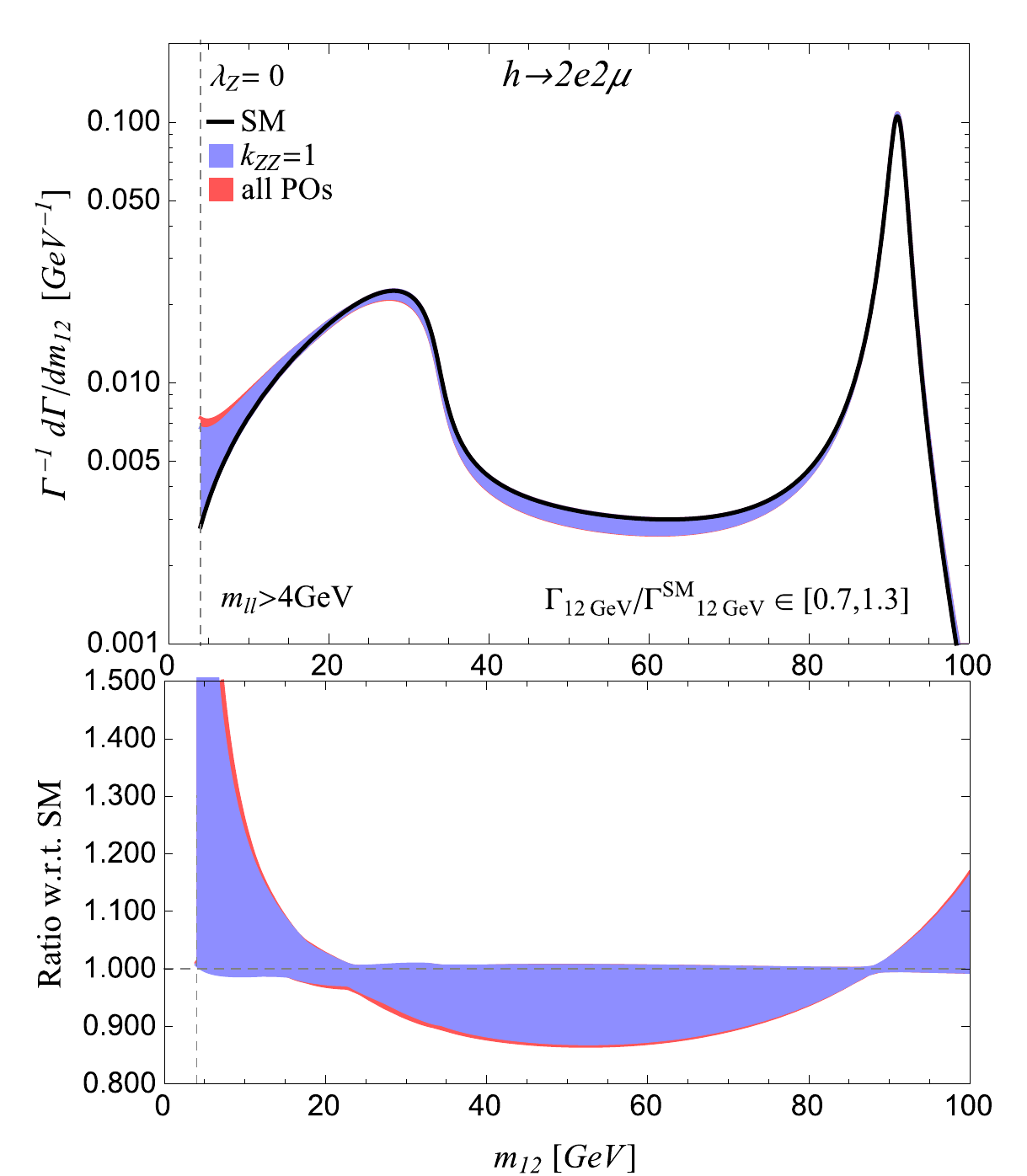} \\
     \end{tabular}
    \caption{\label{fig:Diff_distr_allPOs} \small Single differential distributions in the electron pair invariant mass for the $h\to 2e2\mu$ 
    decay obtained by varying   $\epsilon_{\gamma\gamma,Z\gamma,ZZ}, \epsilon_{Z e_L}$ and $\epsilon_{Z e_R}$ within the $95\%$CL limits from Eqs.~(\ref{eq:gg_Zg_ATLASbounds}-\ref{eq:boundsPOlamZ2}) 
    in the $\lambda_Z \neq 0$ (left) and $\lambda_Z = 0$ (right) case. 
    The blue shaded regions are obtained for $k_{ZZ} = 1$ while the red ones are obtained  using $k_{ZZ} \in [0.5, 1.5]$. In both cases we impose that the total rate, computed with a $12\GeV$ infrared cut on $m_{\ell\ell}$, as in Eq.~\eqref{eq:total_rate_h4l}, is within $30\%$ of the SM one.}
\end{figure*}

In Fig.~\ref{fig:Diff_distr_cont_terms} we present the distributions for $h\rightarrow 2e2\mu$ and $h\rightarrow 4e (4\mu)$, setting $k_{ZZ} = 1$, $\epsilon_{ZZ,Z\gamma,\gamma\gamma} = 0$ and letting vary $\epsilon_{Z e_L}$ and $\epsilon_{Z e_R}$ within their allowed bounds. 
As can be seen, although the effect of the contact terms on the total rate is very large, of $\mathcal{O}(100\%)$ in the $\lambda_Z \neq 0$ case, 
the difference in the shape with respect to the SM is much smaller, namely $\lesssim 15\%$ for $\lambda_Z \neq 0$. A similar cancellation is  present also 
in the $\lambda_Z = 0$ case, although the relative effect is less pronounced. The cancellation of the non-standard effects in the normalized spectrum 
 is, at least in part, a consequence of the strong positive correlation between $\epsilon_{Z e_L}$ and $\epsilon_{Z e_R}$
 shown in  Fig.~\ref{fig:TGCbounds_eps}.

In Fig.~\ref{fig:Diff_distr_eps_ggZgZZ} we study the effect of   $\epsilon_{ZZ,Z\gamma,\gamma\gamma}$  on the invariant-mass distribution. 
Here it is important to notice that the sensitivity to $\epsilon_{Z\gamma,\gamma\gamma}$ depends strongly on the infrared cutoff imposed on the dilepton invariant masses, as expected due to the associated photon pole(s). As shown in Ref.~\cite{Chen:2015iha},
decreasing the cut on $m_{\ell\ell}$
 from $12~{\rm GeV}$   to $4\GeV$ substantially improves the sensitivity to these couplings, even excluding the $m_{\ell\ell}$ region around 
the $\Upsilon$ resonances. 
Moreover, as demonstrated in Ref.~\cite{Gonzalez-Alonso:2014rla}, the irreducible contribution of quarkonium resonances  to the $h\to 4\ell$ spectrum is small and under good theoretical control.

For these reasons, we show the differential decay rate in Fig.~\ref{fig:Diff_distr_eps_ggZgZZ} with a $m_{\ell\ell}^{\rm min}={\rm 4~GeV}$
cut. One observes that
also in this case the effect of the pseudo-observables $\epsilon_{ZZ,Z\gamma,\gamma\gamma}$ on the shape is at most 
of $\mathcal{O}(10\%)$, 
except for the low $m_{\ell\ell}$ region, where the sensitivity is significantly enhanced.
We further stress that lowering the infrared cut on both dilepton invariant masses also leads to an enhancement of the rate 
in the $30-80 \GeV$ region, since then the other lepton pair is allowed to be near the photon pole.

\begin{figure*}[t]
      \centering
    \begin{tabular}{c c}
   \hspace{-0.5cm}
   \includegraphics[width=0.48\textwidth]{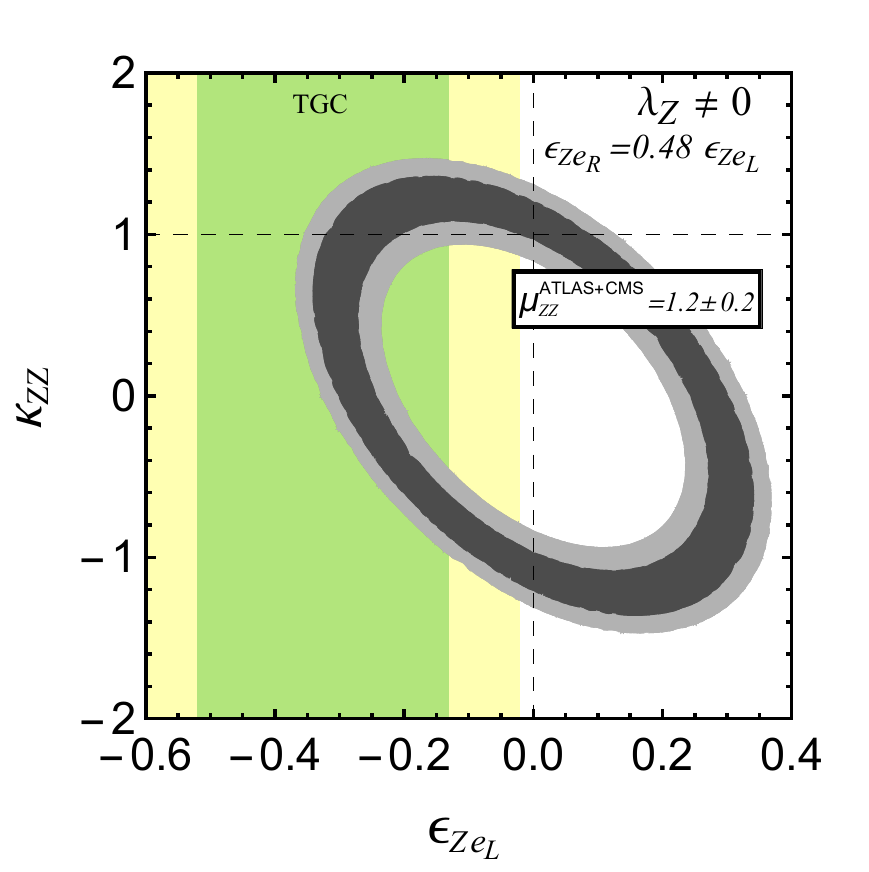} &
      \includegraphics[width=0.48\textwidth]{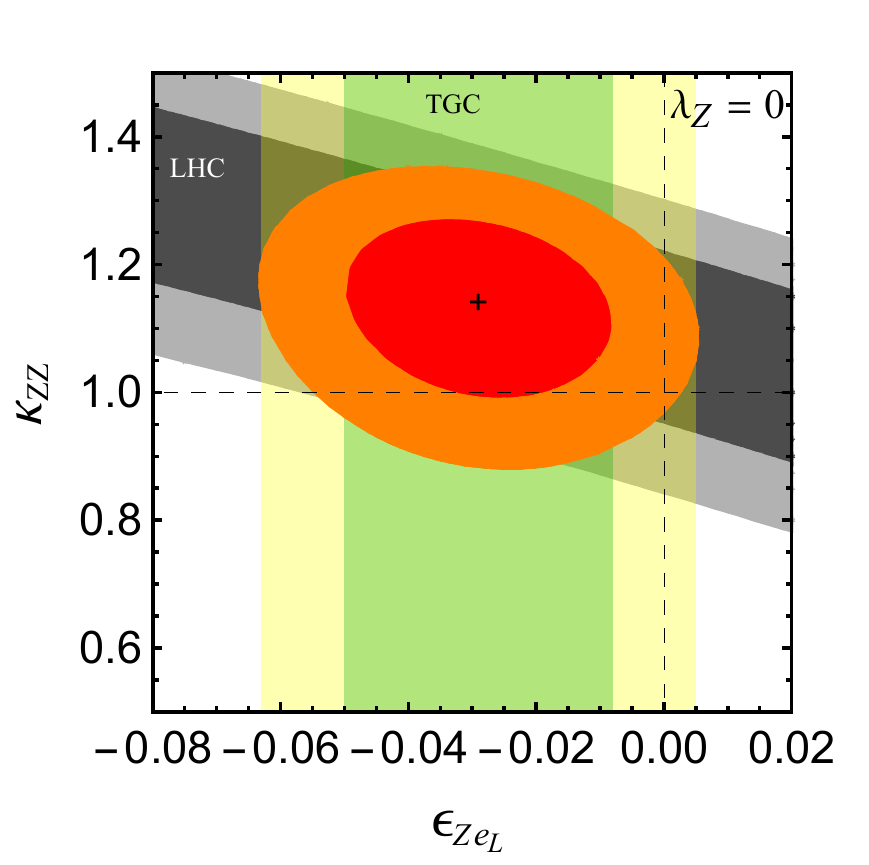}
     \end{tabular}
    \caption{ \small Bounds on  $\kappa_{ZZ}$ vs.~$\epsilon_{Ze_L}$ from the combined ATLAS~\cite{atlas-combination} and CMS~\cite{Khachatryan:2014jba} $h\to 4\ell$ signal strength measurements ($\mu_{ZZ}=1.2\pm0.2$).
    Top: $68\%$ (dark gray) and $95 \%$ (light gray) CL bounds
    valid in the $\lambda_Z\not=0$ case, setting $\epsilon_{Ze_R} = 0.48 \times \epsilon_{Ze_L}$; the
   $68\%$ (yellow) and $95 \%$ (green) CL TGC bounds on $\epsilon_{Ze_L}$ are also shown. 
Bottom: The combined fit of TGC and Higgs data assuming $\lambda_Z=0$, and marginalising 
over $\epsilon_{Ze_R}$, $\epsilon_{ZZ}$ and $\epsilon_{Z\gamma}$,
is shown in red ($68\%$ CL) and orange ($95 \%$ CL). As an illustration, we show in grey the constraint from $\mu_{ZZ}$
measurement only setting $\epsilon_{Ze_R}$, $\epsilon_{ZZ}$ and $\epsilon_{Z\gamma}$ to 0.
\label{fig-kappaZ-epsZe}}
\end{figure*}

Finally, in Fig.~\ref{fig:Diff_distr_allPOs} we show the same plots letting vary all five Higgs PO in Eq.~(\ref{eq:boundsPOlamZ}) within their allowed range. We also impose the total rate to be within $30\%$ of the SM value, as obtained in the global fit of Ref.~\cite{atlas-combination}, analyzing the impact of a 50\% variation in the remaining pseudo-observable 
$\kappa_{ZZ}$ from its SM value.  By construction, this figure summarise the room for non-standard effects in the $h\to4\ell$ shape within the linear EFT scenario. 

The difference between Fig.~\ref{fig:Diff_distr_allPOs}~(left) and ~\ref{fig:Diff_distr_allPOs}~(rigth) could be used, in the future, to indirectly improve the current bounds on $\lambda_Z$.
Notice that $\kappa_{ZZ}$ has a negligible impact on the normalized shape in Fig.~\ref{fig:Diff_distr_allPOs}~(right).
This is because varying $\kappa_{ZZ}$ corresponds to an overall rescaling of the  amplitude assuming a SM-like 
kinematical dependence, and thus this effect cancels at linear order (in $\delta$PO$_i$). 
In the $\lambda_Z \not=0$ case a residual effect survives because 
of quadratic corrections; in the  $\lambda_Z = 0$ case all the PO are constrained to be close to their SM values and 
the effect vanishes, as expected by a consistent EFT  expansion.

\subsection{The $\kappa_{ZZ}$ vs.~$\epsilon_{Z\ell}$ bound from $\Gamma(h\to4 \ell)$.}

From the above discussion we conclude that, even when sizable modifications in the rates occurs, 
namely in the $\lambda_Z \neq 0$ case, the $h\to 4\ell$ spectrum remains SM like  compared to 
the present level of experimental precision. As a result, the only useful bound on the PO that 
can be set at present from $h\to 4\ell$ data, within the linear EFT framework, is the one following from the 
partial decay rates.  We also learned that, given the existing constraints 
on the PO, the structure of  Eq.~\eqref{eq:total_rate_h4l}, and the present level of precision on $\Gamma(h\to 4\ell)$, 
this bound is an effective constraint on $\kappa_{ZZ}$ vs.~$\epsilon_{Z e_L}$, 
fixing $\epsilon_{Z e_R} = 0.48 \times \epsilon_{Z e_L}$, as implied by Fig.~\ref{fig:TGCbounds_eps}.

A detailed extraction of such bound would require a re-analysis of  $h \to 4 \ell$ data. 
However, a good approximation can be obtained as follows: we assume the combined  
result on $\Gamma(h\to 4\ell)/\Gamma_{\rm SM}$  by  ATLAS~\cite{atlas-combination} and CMS~\cite{Khachatryan:2014jba} 
to be $\mu_{ZZ} = 1.2 \pm 0.2$. Using this constraint and setting $\epsilon_{Z e_R}=0.48\times \epsilon_{Z e_L}$ leads to the result in  Fig.~\ref{fig-kappaZ-epsZe} (top).
Assuming the same constraint on  $\Gamma(h\to 4\ell)/\Gamma_{\rm SM}$, setting $\lambda_Z=0$, and allowing all the PO to vary according to Eqs.~\eqref{eq:gg_Zg_ATLASbounds}--\eqref{eq:boundsPOlamZ2}, leads to the result shown 
in  Fig.~\ref{fig-kappaZ-epsZe} (bottom). 
We note that this procedure neglects non-standard effects in $\mu_{ZZ}$ from production and Higgs width, which can be taken into account in a combined re-analysis of Higgs data. In fact, the number chosen here is compatible with
a global fit extraction of $\kappa_Z^2 \sim \Gamma(h\to 4\ell)/\Gamma_{\rm SM}$, see for instance Ref.~\cite{atlas-combination}.

\section{Conclusions}

In this paper we have presented a systematic evaluation 
of the bounds on the  Higgs PO that follow from the EW constraints in the linear EFT regime, 
with particular attention to the PO appearing in $h\to 4\ell$ and $h\to2\ell2\nu$  decays.
Using such bounds  we have derived a series 
 of predictions for $h\to 4\ell$ decay rates and differential distributions. 
 A dedicated analysis of the $h\to 2\ell 2\nu$ modes will be presented 
 in a separate publication.

The results of the EW bounds are summarized in Fig.~\ref{fig:TGCbounds_eps} 
and Eqs.~\eqref{eq:boundsPOlamZ}--\eqref{eq:boundsPOlamZ2}.
We find that, because of the flat direction in TGC bounds ($\lambda_Z \simeq - \delta g_{1,z}$ unconstrained),
EW plus Higgs data leave open the room for sizable $h\to 4\ell$ contact  terms
($\epsilon_{Z\ell}$),  provided they are flavor universal and with the specific 
L--R alignment shown in Fig.~\ref{fig:TGCbounds_eps} (left). 
In principle, $h\to 4\ell$ data can be used to remove the degeneracy in the EFT 
parameter space implied by the TGC flat direction; however, the present level of
precision is not good enough. As a result, the 
uncertainty on the contact terms reflects into a 
 poor knowledge of $\kappa_{ZZ}$, as shown in Fig.~\ref{fig-kappaZ-epsZe} (top).
If the TGC direction is closed (by model-dependent dynamical considerations 
suggesting $\lambda_Z=0$),  the contact terms are bounded at the few percent level,
as shown in Fig.~\ref{fig:TGCbounds_eps} (right),
and have a minor impact in the determination of $\kappa_{ZZ}$,
as shown in Fig.~\ref{fig-kappaZ-epsZe} (bottom).

The phenomenological implications for $h\to 4\ell$ decays of the 
 EW bounds on the Higgs PO are summarized in Fig.~\ref{fig-rates-correlations}--\ref{fig:Diff_distr_eps_ggZgZZ}.
 On the one hand, the uncertainty of the 
 predictions thus obtained determine the level of precision necessary, in future 
 $h\to 4\ell$ analyses,  to improve our constraints on the Higgs linear EFT.
 In this respect, we confirm the conclusion of Ref.~\cite{Pomarol:2013zra,Beneke:2014sba} that shape-modifications in 
 $h\to 4\ell$ are significantly constrained in the linear EFT regime, although we find that deviations from the SM  as large as 
 $10\%$ ($20\%$) are still possible for $\lambda_Z=0$ ($\lambda_Z\not=0$).
 
On the other hand, these predictions  can be interpreted as a series of tests that, if falsified by future $h\to 4\ell$ 
data,  would allow us not only to establish the existence of NP but also to exclude that  $h$ is the massive excitation of a pure $\SU(2)_L$ doublet. In this respect, we stress the firm prediction on the lepton-flavor universality ratios in Fig.~\ref{fig-rates-correlations},
and the bounds on the normalized dilepton invariant-mass spectrum
in Fig.~\ref{fig:Diff_distr_allPOs}.

\begin{acknowledgements}
We thank Adam Falkowski, Francesco Riva, and Michael Trott  for useful comments and discussions.
This research was supported in
part by the Swiss National Science Foundation (SNF) under contract 200021-159720.
M.G.-A. is grateful to the LABEX Lyon Institute of Origins (ANR-10-LABX-0066) of the Universit\'e de Lyon for its financial support within the program ANR-11-IDEX-0007 of the French government.
\end{acknowledgements}

\appendix
\section{Higgs PO, TGC, and the Higgs basis}
\label{app:HiggsBasis}

A dimension-6 operator basis particularly useful to implement experimental constraints on the EFT, at tree-level accuracy, 
is the so-called Higgs basis, developed by the Higgs cross section LHC Working Group \cite{HiggsBasis}.
This basis has been developed following a similar logic to the so-called "BSM primaries" approach of  Ref.~\cite{Gupta:2014rxa,Pomarol:2014dya} (see also Ref.~\cite{Efrati:2015aaa}): the coefficients are specifically built to be directly related to the observables which provide the best constraints on the EFT.   
This basis is defined from an effective Lagrangian describing interactions between mass eigenstates fields with canonical kinetic terms, in the unitary gauge. For the list of the independent couplings and their definition we refer to Ref.~\cite{HiggsBasis}.
The $Z$ and $W$ couplings are defined from the Lagrangian
\ba
	\cL^{d=6}_{Zff} &=& \sum_{f} \sqrt{g^2 + g^{\prime 2}} Z^\mu \bar{f} \delta g^{Zf} \gamma_\mu f ~, \\
	\cL^{d=6}_{Wff'} &=& \frac{g}{\sqrt{2}} W^+_\mu \bar{\nu}_L \gamma^\mu \delta g^{W\ell} e_L + \nonumber \\
	&+& \frac{g}{\sqrt{2}} \left(  W^+_\mu \bar{u}_L \gamma^\mu \delta g^{W q_L} V_{\rm CKM} d_L + W^+_\mu \bar{u}_R \gamma^\mu \delta g^{W q_R} d_R \right)~, \nonumber 
\ea
where $f = e_{L,R},\nu,u_{L,R},d_{L,R}$ and fermion fields all have an implicit flavor index.
For our purposes it is worth stressing that the fermion couplings to $Z$ and $W$ bosons are chosen as independent couplings in this basis. Only the $Z$ couplings to neutrinos and the $W$ couplings to left-handed quarks are dependent of the others:
\be
	\delta g^{Z \nu} = \delta g^{Z e_L} + \delta g^{W \ell}$~, \quad $\delta g^{W q_L} = \delta g^{Z u_L} - \delta g^{Z d_L}~.
	\label{eq:ZWcouplEFTRel}
\ee
This implies that, at tree-level, SLD and LEP-I pseudo-ob\-ser\-vables from the $Z$-pole  and from $W$ decays can be directly related to these couplings. In particular deviations in the $Z$ couplings can be constrained at the per-mil level while deviations in the $W$ couplings only at the percent level \cite{Efrati:2015aaa}.

The tree-level matching between the Higgs pseudo-ob\-ser\-vables of Ref.~\cite{Gonzalez-Alonso:2014eva} and the EFT coefficients in the Higgs basis  is
\ba
	\kappa_{ZZ} &=& 1 + \delta c_z + g^2 c_{z \Box}~,\nonumber \\
	\kappa_{WW} &=& 1 + \delta c_w + g^2 c_{w \Box}~,\nonumber \\
	\epsilon_{Zf} &=&\frac{2 m_Z}{v} \left( \delta g^{Zf} + \frac{g^2}{2} (T_3^f - Q_f s_\theta^2)  c_{z \Box} + \frac{e^2 Q_f}{2} c_{\gamma\Box} \right),\nonumber \\
	\epsilon_{Wf} &=& \frac{\sqrt{2} m_W}{v} \left( \delta g^{Wf} + \frac{g^2}{2} c_{w \Box} \right)~,\nonumber \\
	\epsilon_{ZZ} &=& \epsilon_{ZZ}^{\rm SM-1L} - \frac{g^2 + g^{\prime 2}}{2} c_{zz}~, \\
	\epsilon_{Z\gamma} &=& \epsilon_{Z\gamma}^{\rm SM-1L} - \frac{g g^{\prime}}{2} c_{z\gamma} ~,\nonumber \\
	\epsilon_{\gamma\gamma} &=& \epsilon_{\gamma\gamma}^{\rm SM-1L} - \frac{e^2}{2} c_{\gamma\gamma} ~,\nonumber \\
	\epsilon_{WW} &=& \epsilon_{WW}^{\rm SM-1L} - \frac{g^2}{2} c_{ww}  ~, \nonumber 
	\label{eq:HPOinHiggsBasis}
\ea
where the dependent couplings $\delta c_w, c_{ww}, c_{w\Box}, c_{\gamma\Box}$ are given, in terms of the independent ones, by \cite{HiggsBasis}
\ba
	\delta c_w	&=& \delta c_z + 4 \delta m~, \nonumber \\
	c_{ww} 	&=& c_{zz} + 2 s_\theta^2 c_{z\gamma} + s_\theta^4 c_{\gamma\gamma}~, \\
	c_{w \Box}&=& \frac{1}{g^2 - g^{\prime 2}} \left[ g^2 c_{z\Box} + g^{\prime 2} c_{zz} - e^2 s_w^2 c_{\gamma\gamma} - (g^2 - g^{\prime 2} ) s_w^2 c_{z\gamma} \right]~,\nonumber \\
	c_{\gamma \Box} &=& \frac{1}{g^2 - g^{\prime 2}} \left[ 2 g^2 c_{z\Box} + (g^2 + g^{\prime 2}) c_{zz} - e^2 c_{\gamma\gamma} - (g^2 - g^{\prime 2} ) c_{z\gamma} \right]~. \nonumber 
\ea
We  stress here that the choice of keeping $c_{z\Box}$ as an independent coupling instead of $\delta c_w$ implies that the pseudo-observables $\kappa_{ZZ}$ and $\kappa_{WW}$ are not in one-to-one correspondence with the $\delta c_z$ and $\delta c_w$ couplings, and also the contact terms $\epsilon_{Vf}$ are not in one-to-one correspondence with the $c^{Vf}$ coefficients defined in Ref.~\cite{HiggsBasis}. Even though this choice is not optimal for our purposes, the relations between observables presented in Sec.~\ref{sec:LEPandHPO} are of course independent on this basis choice.

The CP-conserving anomalous TGC are defined by the Lagrangian
\ba
	\cL^{\rm TGC} &=& i e \delta \kappa_\gamma A_{\mu\nu} W^{+\mu} W^{- \nu} + i g c_\theta \delta \kappa_z Z^{\mu\nu} W^+_\mu W^-_\nu \nonumber \\
		&+& i g c_\theta \delta g_{1,z} (W^+_{\mu\nu} W^{-\mu} - W^-_{\mu\nu} W^{+\mu})Z^\nu + \\
		&+& i \frac{g c_\theta}{m_W^2} \lambda_{Z} W^+_{\mu\nu} W^{-\nu\rho} Z_\rho^\mu + i \frac{e}{m_W^2} \lambda_{\gamma} W^+_{\mu\nu} W^{-\nu\rho} A_\rho^\mu~. \nonumber 
\ea
In general, at dimension-6 in the linear EFT, $\delta \kappa_z = \delta g_{1,z} - t_\theta^2 \delta \kappa_\gamma$ and $\lambda_\gamma = \lambda_Z$. Moreover, in this basis also $\delta g_{1,z}$ and $\delta \kappa_\gamma$ are dependent couplings:
\ba
	\delta g_{1,z} &=& \frac{1}{2(g^2 - g^{\prime 2})} \left[ - g^2 (g^2 + g^{\prime 2}) c_{z\Box} - g^{\prime 2} (g^2 + g^{\prime 2}) c_{zz} + \right. \nonumber \\
	&& + \left. e^2 g^{\prime 2} c_{\gamma\gamma} +g^{\prime 2} (g^2 - g^{\prime 2} ) c_{z\gamma} \right]~, \label{eq:TGCdependent} \\
	\delta \kappa_\gamma &=& - \frac{g^2}{2} \left( c_{\gamma\gamma} \frac{e^2}{g^2 + g^{\prime 2}} + c_{z\gamma} \frac{g^2 - g^{\prime 2}}{g^2 + g^{\prime 2}} - c_{zz} \right)~, \nonumber 
\ea
while only $\lambda_Z$ is an independent coupling. Since we are interested in studying the constraints from TGC on Higgs observables it is convenient for us to exchange the two independent Higgs couplings $c_{z\Box}$ and $c_{zz}$ in favour of these TGC using Eq.~\eqref{eq:TGCdependent}. By doing so and substituting the result in Eq.~\eqref{eq:HPOinHiggsBasis} we obtain the relations 
of Sec.~\ref{sec:LEPandHPO}. We also checked independently those relations by employing a basis 
of manifestly $\SU(2)_L\times\U(1)_Y$ invariant operators. 

Once the per-mil constraints from LEP-1 measurements have been imposed, the TGC can be constrained from a fit to LEP-2 
data on $\sigma(e^+e^-\to WW)$ and single $W$ production. We use the results of the fit performed in Ref.~\cite{Falkowski:2014tna}:
\ba
	\left( \begin{array}{c}
		\delta g_{1Z} \\
		\delta \kappa_{\gamma} \\
		\lambda_Z
	\end{array}\right) &=& 
	\left( \begin{array}{c}
		-0.83 \pm 0.34 \\
		0.14 \pm 0.05 \\
		0.86 \pm 0.38
	\end{array}\right)~, \nonumber \\
	\rho &=& \left( \begin{array}{ccc}
		1 & - 0.71 & -0.997 \\
		.  & 1  & 0.69 \\
		. & . & 1
	\end{array}\right)~.
\ea
The large allowed range for $\delta g_{1,z}$ and $\lambda_Z$ is due to an approximately blind direction in LEP-2 $WW$ data corresponding to $\lambda_Z \approx - \delta g_{1,z}$ \cite{Brooijmans:2014eja}. 
This implies a very loose bound on $\delta g_{1,z}$ upon marginalizing on $\lambda_Z$. Since in a wide class of explicit 
ultraviolet completions of the linear EFT 
$\lambda_Z$ is expected to be loop suppressed compared to $\delta g_{1,z}$ and $\delta \kappa_\gamma$, 
it is worth considering explicitly the case $\lambda_Z = 0$. In this limit the bound on $\delta g_{1,z}$ is much stronger:
\be
	\left( \begin{array}{c}
		\delta g_{1Z} \\
		\delta \kappa_{\gamma} \\
	\end{array}\right) = 
	\left( \begin{array}{c}
		-0.06 \pm 0.03 \\
		0.06 \pm 0.04
	\end{array}\right)~, \quad
	\rho = \left( \begin{array}{cc}
		1 & - 0.5 \\
		.  & 1 
	\end{array}\right)~.
	\label{eq:tgc-O}
\ee

The $W$ couplings to electron and muon are constrained at the percent level. We use the results from the non-universal fit of Ref.~\cite{Efrati:2015aaa}:
\be\begin{split}
	\left( \begin{array}{c}
		\delta g^{We} \\
		\delta g^{W\mu} \\
		\delta g^{W\tau}
	\end{array}\right) &= 
	\left( \begin{array}{c}
		-1.00 \pm 0.64 \\
		-1.36 \pm 0.59 \\
		1.95 \pm 0.79
	\end{array}\right)
	\times 10^{-2}~, \\
	\rho &= \left( \begin{array}{ccc}
		1 & - 0.12 & -0.63 \\
		.  & 1  & -0.56 \\
		.  & .  & 1
	\end{array}\right)~.
	\label{eq:WCouplLept}
\end{split}\ee
These bounds can be used to constrain at the percent level the $Z$ couplings to neutrinos, using Eq.~\eqref{eq:ZWcouplEFTRel}.


\end{document}